\documentclass[preprint,floatfix,amsmath,amssymb,superscriptaddress,longbibliography]{revtex4-1}

\usepackage{graphicx}
\usepackage{bm}
\usepackage{xr}
\usepackage{float}
\usepackage{lineno}
\usepackage{siunitx}
\DeclareSIUnit\Oersted{Oe}
\usepackage{xcolor}
\begin{document}

%Title of paper
\title{The spin Hall effect of Bi-Sb alloys driven by thermally excited Dirac-like electrons}

% author & affiliation
\author{Zhendong Chi}
\affiliation{Department of Physics, The University of Tokyo, Bunkyo-ku, Tokyo 113-0033, Japan}
\affiliation{National Institute for Materials Science, Tsukuba, Ibaraki 305-0047, Japan}
\author{Yong-Chang Lau}
\email{yongchang.lau@qspin.phys.s.u-tokyo.ac.jp}
\affiliation{Department of Physics, The University of Tokyo, Bunkyo-ku, Tokyo 113-0033, Japan}
\affiliation{National Institute for Materials Science, Tsukuba, Ibaraki 305-0047, Japan}
\author{Xiandong Xu}
\affiliation{National Institute for Materials Science, Tsukuba, Ibaraki 305-0047, Japan}
\author{Tadakatsu Ohkubo}
\affiliation{National Institute for Materials Science, Tsukuba, Ibaraki 305-0047, Japan}
\author{Kazuhiro Hono}
\affiliation{National Institute for Materials Science, Tsukuba, Ibaraki 305-0047, Japan}
\author{Masamitsu Hayashi}
\email{hayashi@phys.s.u-tokyo.ac.jp}
\affiliation{Department of Physics, The University of Tokyo, Bunkyo-ku, Tokyo 113-0033, Japan}
\affiliation{National Institute for Materials Science, Tsukuba, Ibaraki 305-0047, Japan}

\date{\today}

% abstract
\begin{abstract}
We have studied the charge to spin conversion in Bi$_{1-x}$Sb$_x$/CoFeB heterostructures. The spin Hall conductivity (SHC) of the sputter deposited heterostructures exhibits a high plateau at Bi-rich compositions, corresponding to the topological insulator phase, followed by a decrease of SHC for Sb-richer alloys, in agreement with the calculated intrinsic spin Hall effect of Bi$_{1-x}$Sb$_x$ alloy. The SHC increases with increasing thickness of the Bi$_{1-x}$Sb$_x$ alloy before it saturates, indicating that it is the bulk of the alloy that predominantly contributes to the generation of spin current; the topological surface states, if present in the films, play little role.
Surprisingly, the SHC is found to increase with increasing temperature, following the trend of carrier density.
%The equivalent spin current mobility, which is defined as the SHC divided by the carrier density, is an order of magnitude larger for the Bi-rich Bi$_{1-x}$Sb$_x$ compared to typical transition metals.
These results suggest that the large SHC at room temperature, with a spin Hall efficiency exceeding 1 and an extremely large spin current mobility, is due to increased number of Dirac-like, thermally-excited electrons in the $L$ valley of the narrow gap Bi$_{1-x}$Sb$_x$ alloy.
\end{abstract}

% PACS numbers

%\maketitle
\maketitle

%\linenumbers
%main
\section*{INTRODUCTION}
Generation of spin current or flow of spin angular momentum lies at the heart of modern spintronics. The power consumption of a spintronic device is directly related to its efficiency for converting a charge current that dissipates energy to a dissipative spin polarized current or a dissipationless spin current\cite{Murakami2003science,LeeScience2004}.
A conventional means for creating a flow of spin angular momentum is by passing a charge current across a ferromagnetic metal (FM) that converts to a spin polarized current. The efficacy of this process is proportional to the spin polarization of the FM.
More recently, generation of spin current from a charge current passed along a non-magnetic metal (NM)\cite{DyakonovZPR1971,HirschPRL1999,SinovaRMP_SHE_review} or interface of materials with strong spin orbit coupling\cite{BychkovJETP1984,Edelstein1990SSC,Manchon2015} has emerged as an attractive alternative.
%The spin Hall effect (SHE) \cite{DyakonovZPR1971,HirschPRL1999,SinovaRMP_SHE_review,HoffmannSHE_Review} and/or other spin-orbit phenomena \cite{BychkovJETP1984,Manchon2015} are exploited as means to generating a flow of spin angular momentum.
In particular, the discovery\cite{LiuScience2012_SHE_Ta} of the giant spin Hall effect (SHE) in $5\textit{d}$ transition heavy metals (HM) have triggered significant effort in exploiting the spin current to electrically control magnetization of ferromagnets placed nearby.
In HM/FM bilayer systems, the magnetization of the FM layer can absorb the orthogonal component of the non-equilibrium spin density originating from the SHE, giving rise to current-induced spin-orbit torque (SOT) at the HM/FM interface \cite{Miron2010nmat,KimNatMat2012_Harmonic,GarelloNatNano2013_SOT}.
%It is now well established that the spin Hall effect (SHE) \cite{DyakonovZPR1971,HirschPRL1999,SinovaRMP_SHE_review,HoffmannSHE_Review} and other spin-orbit phenomena, e.g. the Rashba-Edelstein effect \cite{BychkovJETP1984,Manchon2015}, can provide sufficient spin angular momentum that can be exploited to manipulate magnetization of the adjacent FM layer.
%The charge-to-spin conversion process of the Rashba-Edelstein effect relies on the interface electronic state, often parametrized using the so-called Edelstein length, whereas the SHE is governed by the spin Hall conductivity (SHC) of the paramagnet.
The SOT in such bilayers enabled current induced magnetization switching\cite{MironNature2011,LiuScience2012_SHE_Ta}, current driven motion of chiral domain walls and skyrmions\cite{Emori2013,Ryu2013,Woo2016nmat}, and magnetoresistance effect that depends on the SHE, often referred to as the spin Hall magnetoresistance\cite{Nakayama2013PRL,Chen2013PRB}.
The figure of merit of the charge to spin conversion in SHE is known as the damping-like spin Hall efficiency $\xi_\textrm{DL}$ that includes non-ideal spin transmission across the interface\cite{RojasSanchez2014PRL,Zhang2015nphys}.
Using $\xi_\textrm{DL}$, the spin current $j_\textrm{s}$ generated from a charge current $j_\textrm{c}$ passed to a  non-magnetic metal layer and entering the FM layer can be expressed as $j_\textrm{s}=\xi_\textrm{DL}(\hbar/2e)j_\textrm{c}$ where $\hbar$ is the reduced Planck constant and $e$ is the electrical charge.
%$\xi_\textrm{DL}$ is the effective damping-like spin Hall efficiency including non-ideal spin transmission across the interfaces.
As $\xi_\textrm{DL}$ may depend on the longitudinal conductivity $\sigma_\textrm{xx}$ of the NM layer which varies with extrinsic factors such as impurity concentration and film texture, it is customary to use the spin Hall conductivity (SHC) $\sigma_\textrm{SH}$, defined through the relation $\sigma_\textrm{SH} = \xi_\textrm{DL} \cdot \sigma_\textrm{xx}$, to provide a measure of the anomalous transverse velocity the carriers obtain via the SHE\cite{SinovaRMP_SHE_review}.

Recent advances in the understanding of topological insulators\cite{Hasan2010RMP} and Weyl semimetals\cite{Sun2016PRL} have attracted great interest in exploiting their unique electronic states for spintronic applications.
Giant charge to spin conversion efficiencies were found in heterostructures that consist of a topological insulator and a ferromagnetic/ferrimagnetic layer\cite{MellnikNature2014,Fan2014nmat,JamaliNanoLetter2015,KondouNatPhys2016,Yasuda_PRL2017_Transverse_scattering_TI,Wang2017,Han2017PRL,MahendraNatMat2018_BiSe_sputter_SOT,NguyenNatMat2018_BiSb_MBE,Li2018SciAdv}.
The large charge to spin conversion efficiency observed in such systems were attributed to the current induced generation of spin density enabled by the spin momentum locked surface states of topological insulators.
Ideally the bulk of a topological insulator should be insulating. In practice, however, it remains as a great challenge to limit the current flow within the bulk of this material class. This is particularly the case for thin film heterostructures in which imperfect crystal structures and interdiffusion with the adjacent layers may reduce or eliminate the band gap of the bulk state. To take advantage of the topological surface states in generating spin accumulation, it has been considered detrimental to have current paths in the bulk.
In terms of bulk conduction of carriers, the charge to spin conversion efficiency of Bi, a small gap semimetal with large spin orbit coupling\cite{Liu1995PRB,Teo2008PRB} and being one of the most used elements in forming topological insulators, has been reported to be extremely small\cite{Hou2012APL,EmotoPRB2016} compared to the 5\textit{d} transition metals.
Theoretically, Bi and Bi-Sb alloys have been predicted\cite{Fuseya2012jpsj,SahinPRL2015_BiSb_SHC,Fukuawa2017jpsj} to exhibit considerable SHC due to its unique electronic state.

Here, we show that the charge to spin conversion efficiency that originates from the bulk of Bi$_{1-x}$Sb$_x$ alloys is significantly larger than that of the 5\textit{d} transition metals. The SHC of the alloy increases with increasing thickness before it saturates. Such thickness dependence of the SHC, together with its facet independence, suggest that a significant amount of spin current is generated from the bulk of the alloy: we find little evidence of spin current generation from the topological surface states, if they were to exist in the sputtered films used here. The damping-like spin Hall efficiency exceeds 1 for the Bi-rich Bi$_{1-x}$Sb$_x$ alloy and decreases with increasing Sb concentration.
% being one of the largest values reported for SHE originating from the bulk of the film. 
The alloy composition dependence of the SHC indicates that the SHE of the alloy has considerable contribution from the so-called intrinsic SHE. Surprisingly, we find that the SHC and the spin Hall efficiency increases with increasing measurement temperature. We find up to three-fold (two-fold) enhancement of $\sigma_\textrm{SH}$ ($\xi_\textrm{DL}$) upon increasing the temperature from $\SI{10}{\kelvin}$ to $\SI{300}{\kelvin}$. Although thermal fluctuation is typically detrimental for many key parameters of spintronic devices, the SHC of Bi$_{1-x}$Sb$_x$ alloy is enhanced at higher temperature due to the increased number of carriers at the valleys with large Berry curvature. Our results suggest that such carriers in Bi$_{1-x}$Sb$_x$ alloys, particularly in the Bi-rich compositions, possess large spin current generation efficiency and equivalent spin current mobility.

\section*{RESULTS}
\subsection*{Structural characterizations}
Thin film heterostructures with base structure of Sub./seed/[$t_{\textrm{Bi}}$ Bi$\vert$$t_{\textrm{Sb}}$ Sb]$_{N}$/$t_{\textrm{Bi}}$ Bi/{\textit t$_\textrm{CoFeB}$} CoFeB/2 MgO/1 Ta (thicknesses in nanometer) were grown by magnetron sputtering at ambient temperature on thermally oxidized Si substrates. $\textit{N}$ represents the number of repeats of the Bi$\vert$Sb bilayers.
%use [Bi$\vert$Sb] in short for the repeated layers and
The thicknesses of the Bi ($t_{\textrm{Bi}}$) and Sb ($t_{\textrm{Sb}}$) layers in the repeated structure is set to meet a condition $t_{\textrm{Bi}} + t_{\textrm{Sb}}$$\sim$0.7 nm.
%i.e. their sum equals roughly 1 quintuple layer (QL) of typical Bi-Sb alloy.
%\textcolor{blue}{(The structure of BiSb is different from Bi2Te3. The quintuple description should not be appropriate for BiSb.)}
%, which corresponds to $t_{\textrm{BiSb}}$ ranging from 3.0 to 16.2 nm. hereafter using nominal thicknesses {\textit t$_\textrm{BiSb}$} [Bi$\vert$Sb] in short for [0.34 Bi$\vert$0.32 Sb]$_{N}$)
%Hereafter, we label these structures as BS4, BS6, BS8, BS12, BS16, BS20, and BS24, respectively.
%Two series of flat films of varying \textit t$_\textrm{CoFeB}$ and with $N=8$ ($t_{\textrm{BiSb}}=\SI{5.6}{\nano\meter}$) or $N=16$ ($t_{\textrm{BiSb}}=\SI{10.9}{\nano\meter}$) were grown to study the \textit{N}-dependence, if any, of the structural and magnetic properties.
The nominal composition of CoFeB is Co:Fe:B=20:60:20 at\%.
Unless noted otherwise, we use 0.5 nm Ta as the seed layer for Bi$\vert$Sb multilayers. The capping layer is always fixed to 2 MgO/1 Ta. We assume the top 1 nm Ta layer is fully oxidized and does not contribute to the transport properties of the films.

\begin{figure}
\centering
\includegraphics[width=1\columnwidth]{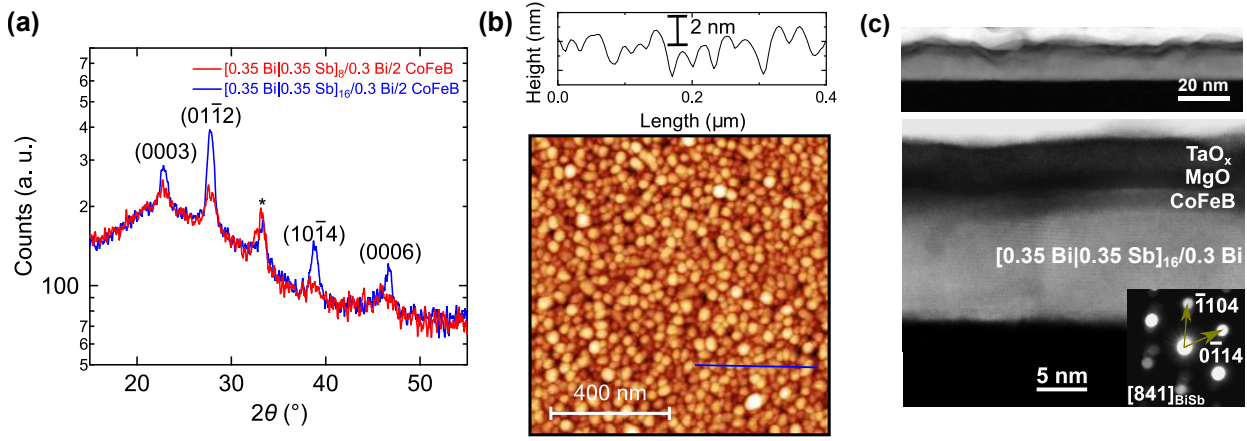}\\
\caption{\textbf{Structural characterization of Bi$\vert$Sb multilayers.} (\textbf{a}) X-ray diffraction (XRD) spectra of 0.5 Ta/[0.35 Bi$\vert$0.35 Sb]$_{N}$/0.3 Bi/2 CoFeB/2 MgO/1 Ta with $N=8$ (blue line) and $N=16$ (red line). %\textcolor{red}{(Index the peak at 32 deg or comment on it?)}\textcolor{blue}{(Done)}
(\textbf{b}) AFM image of the $N=8$ film. A line profile along the blue solid line drawn in the bottom image is shown at the top panel.
(\textbf{c}) Cross-sectional HAADF-STEM images of the $N=16$ structure. %\textcolor{red}{Change [Bi$\vert$Sb] in the TEM image to [0.3 Bi$\vert$0.3 Sb]$_{16}$/0.3 Bi. The same for the XRD data.}\textcolor{blue}{(Done)}
Selected nanobeam diffraction pattern of the Bi$\vert$Sb multilayer is shown in the inset.
%\textcolor{red}{(The font size of (a) is too big, pls reduce.)}\textcolor{blue}{(Done)}\textcolor{red}{Sorry, I meant the text size of the figure axis, not the text inside. Reduce the font size of the axis title and numbers, increase the size for the inner texts.}\textcolor{blue}{(Done)}
}
\label{fig:Fig_1}
\end{figure}

$\theta - 2\theta$ X-ray diffraction (XRD) spectra of representative films with $t_{\textrm{Bi}}$$\sim$ $t_{\textrm{Sb}}$$\sim$ 0.35 nm, $N=8,16$ are shown in Fig.~\ref{fig:Fig_1}(a).
%XRD spectra of 10 Bi/2 CoFeB and 10 Sb/2 CoFeB films are also shown in the same plot for comparison.
The films are polycrystalline and the peaks are indexed based on the hexagonal representation of the rhombohedral Bi$_{1-x}$Sb$_x$ (space group R$\bar3$m; No. 166) that forms solid solution throughout the composition.
%The preferred orientation along the normal of the Sb (Bi) film is the (0003)((01$\bar1$2) and (10$\bar1$4)) direction.
Bragg diffraction peaks corresponding to (0003), (01$\bar1$2) and (10$\bar1$4) crystallographic directions are found. The peak intensities increase with increasing \textit{N}, reflecting improved crystallinity of the film.
%The peak positions of Bi$_{0.53}$Sb$_{0.47}$ lying between those of Bi and Sb indicate that the lattice constants of the alloy are intermediate between those of the two end compounds, as expected from the Vegard's law.
Atomic force microscopy (AFM) image of the $N=8$ film is shown in Fig.~\ref{fig:Fig_1}(b). The root mean square (r.m.s.) roughness is of the order of 1 nm. %The roughness is a little larger than the granular Bi$_x$Se$_{1-x}$ films grown by magnetron sputtering from a BiSe alloy target\cite{MahendraNatMat2018_BiSe_sputter_SOT}.
Representative cross sectional high-angle annular dark-field scanning transmission electron microscopy (HAADF-STEM) images of the $N=16$ film are shown in Fig.~\ref{fig:Fig_1}(c). The lower magnification STEM image at the upper panel confirms that the Bi$\vert$Sb multilayer is granular and continuous. % over a length of a fraction of micrometer.
We find the average grain size is $\sim$$\SI{35}{\nano\meter}$. The 2 nm CoFeB and the subsequent capping layers are also continuous and follow the morphology of the multilayer. The lower panel shows the high-resolution STEM image of the film. The lattice fringes clearly seen in the image reveals the good crystallinity of Bi$\vert$Sb multilayer. A typical nanobeam diffraction pattern of the Bi$\vert$Sb multilayer is shown in the inset of Fig.~\ref{fig:Fig_1}(c). The diffraction patterns suggest that the grains are consisted of Bi$_{1-x}$Sb$_{x}$ nanocrystallites with random orientations within the film plane. Although alternating Bi and Sb layers were sputtered to form Bi$\vert$Sb multilayers,
%we do not find superlattice diffraction spots associated with the repeated structure. This is supported by
energy-dispersive X-ray spectroscopy (EDS) mapping (see supplementary material) show that the two elements intermix to form an alloy rather than a layer-by-layer superlattice.
%We observe opposite concentration gradient of the two elements along the $z$-axis, whereby a higher Bi (Sb) content is found at the bottom (upper) interface with Ta (CoFeB). This may be attributed to the finite solubility of Sb in Co and Fe, in contrast to Bi which is practically immiscible with any of these elements.

\subsection*{Experimental setup}
\begin{figure}
\centering
\includegraphics[width=0.85\columnwidth]{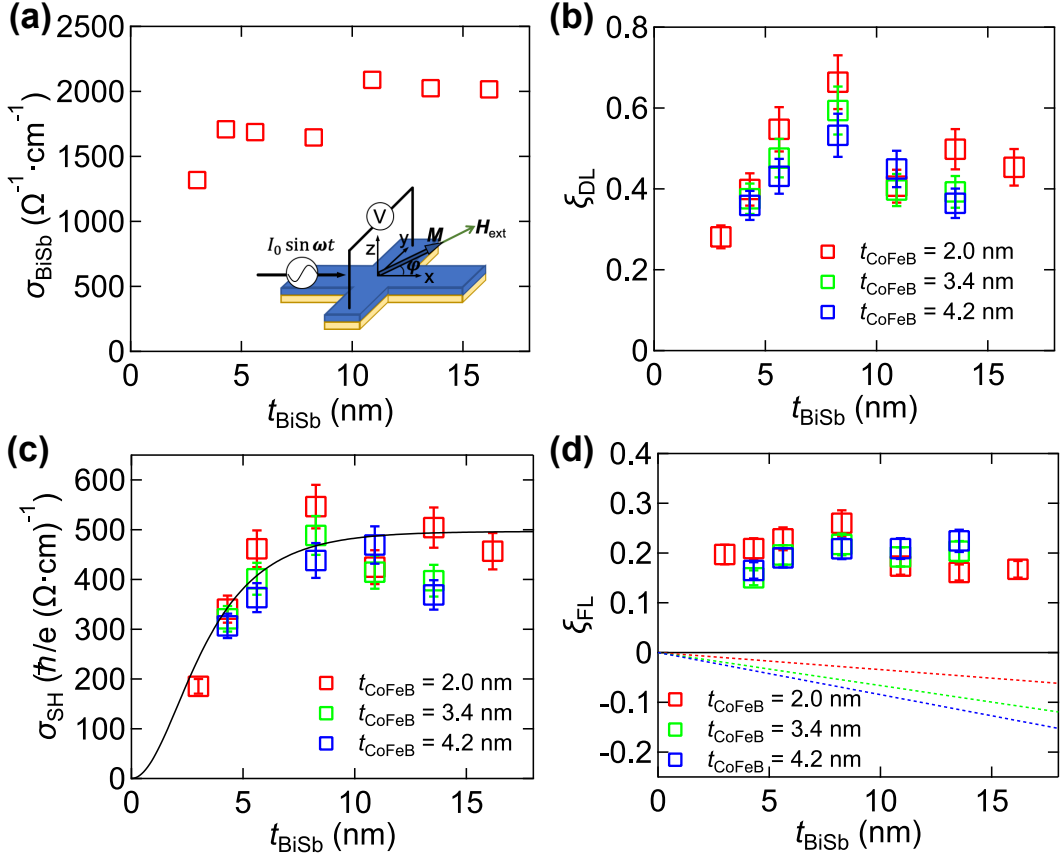}\\
\caption{\textbf{Bi$_{0.53}$Sb$_{0.47}$ thickness dependence of SOT and $\sigma_\textrm{SH}$.} (\textbf{a}) The conductivity $\sigma_\textrm{BiSb}$ of Bi$_{0.53}$Sb$_{0.47}$ plotted as a function of its thickness $t_{\textrm{BiSb}}$. Inset: schematic illustration of a Hallbar device and the coordinate system. (\textbf{b}-\textbf{d}) $t_{\textrm{BiSb}}$ dependence of the damping-like spin Hall efficiency $\xi_\textrm{DL}$ (\textbf{b}), the spin Hall conductivity $\sigma_\textrm{SH}$ (\textbf{c}) and the field-like spin Hall efficiency $\xi_\textrm{FL}$ (\textbf{d}) of $t_{\rm{BiSb}}$ Bi$_{0.53}$Sb$_{0.47}$/$t_{\rm{CoFeB}}$ CoFeB. Dotted lines represent contributions from the Oersted field. All data were obtained at \SI{300}{\kelvin}.
%\textcolor{red}{(Use the same unit expression for (a) and (c)? (c) looks better.)}\textcolor{blue}{(Done)}
}
\label{fig:Fig_2}
\end{figure}

Since structural characterization show that the two elements intermix and form an alloy, we denote, hereafter, the Bi$\vert$Sb multilayers (i.e. [$t_{\textrm{Bi}}$ Bi$\vert$$t_{\textrm{Sb}}$ Sb]$_{N}$/$t_{\textrm{Bi}}$ Bi)
as $t_{\textrm{BiSb}}$ Bi$_{1-x}$Sb$_{x}$ using the total thickness of the multilayer ($t_{\textrm{BiSb}}$) and the corresponding composition $x$ defined by the relative thickness of the Bi and Sb layers, i.e. $x \equiv \frac{t_{\textrm{Sb}}}{t_{\textrm{Bi}} + t_{\textrm{Sb}}}$.
%\textcolor{blue}{(Definition of $x$ has been modified.)}
To evaluate the SOT, we pattern Hall bar devices using optical lithography.  The nominal channel width $w$ and length $L$ are set to $\SI{10}{\micro\meter}$ and $\SI{25}{\micro\meter}$, respectively.
%\textcolor{red}{(L is 25 um for the standard Hall bar. Please correct the resistivity if you have used 20 um)}\textcolor{blue}{(The length of the wire should be 50 um. We find the resistance of the whole wire is about 2.5 times of that measured between two Hall contacts. Thus we consider $L$ should be 20 um in our samples; $L$ may be defined by the distance of the inner edges of the two Hall arms and not by the center of the two arms.)}\textcolor{red}{I don't understand, the length is defined by the mask design. I think the typical Hall bar length between the two voltage probes (from center to center) is 25 um.}\textcolor{blue}{(Based on the resistance measurement, we think the inner-edge-to-inner-edge length (20um from the mask design) may be more appropriate for defining $L$)}
Illustration of the Hall bar device and the coordinate system adapted in this work are schematically illustrated in the inset of Fig.~\ref{fig:Fig_2}(a).
The longitudinal resistance $R_\textrm{xx}$ and the transverse resistance $R_\textrm{xy}$ of the devices were obtained using direct current (DC) transport measurements. Linear fitting to the sheet conductance $L/(w R_\textrm{xx})$ versus the thickness of one of the layers
%\textcolor{red}{(Is the BiSb conductivity obtained from the CoFeB wedge films? The y-intercept is usually not that accurate.)}\textcolor{blue}{(Yes, we chose this method. The slope of $t_{\rm{CoFeB}}$ vs sheet resistance gives a good value of $\sigma_{\rm{CoFeB}}$ close to 170 $\mu\Omega\cdot\rm{cm}$, so we also think the $\sigma_{\rm{BiSb}}$ is right.)}\textcolor{red}{But you also have the BiSb wedge, do the numbers agree?}\textcolor{blue}{(10nm Bi0.53Sb0.47 double wedge gives $\rho_{\rm{BiSb}} \sim$ 690 uOhm cm considering $\rho_{\rm{CoFeB}}$ is 150 uOhm cm. $\rho_{\rm{BiSb}}$ for 10nm film sample is around 600 um cm, a bit smaller than the wedge one. But we note that the resistance of the devices changes (by about 10 percent) over time, older samples tend to become more resistive. All these variations are within the symbol size of the graph and will not alter the main conclusion)}
is used to estimate the conductivity $\sigma_\textrm{X}$ (X=BiSb, CoFeB, seed layer). The current distribution within the heterostructures is calculated using the thicknesses and conductivities of the conducting layers.

We use the harmonic Hall technique \cite{KimNatMat2012_Harmonic,GarelloNatNano2013_SOT,Kawaguchi_2013,AvciPRB2014_R2w_thermoelectric,LauJJAP2017_Harmonic_IP} to quantify the SOT of in-plane magnetized Bi$_{1-x}$Sb$_{x}$/CoFeB heterostructures. Upon applying an alternating current $I_0 \sin{\omega t}$ with frequency $\omega$ and amplitude $I_0$ along $\bm{x}$, an external magnetic field $\bm{H}_{\rm{ext}}$ %(amplitude: $H_{\rm{ext}}$)
is applied in the \textit{xy} plane while the in-phase first harmonic $(V_{1\omega})$ and the out-of-phase second harmonic $(V_{2\omega})$ Hall voltages along $\bm{y}$ are simultaneously measured. The Hall resistance is obtained by dividing the harmonic voltages with $I_0$, \textit{i.e.} $R_{1\omega(2\omega)} \equiv V_{1\omega(2\omega)} / I_0$. $R_{1\omega}$ is dominated by the planar Hall and anomalous Hall resistances, whereas $R_{2\omega}$ contains contributions from current-induced damping-like spin-orbit effective field ($H_{\rm{DL}}$), the field-like spin-orbit effective field ($H_{\rm{FL}}$), the Oersted field ($H_{\rm{Oe}}$) and thermoelectric effects (anomalous Nernst effect (ANE) of CoFeB, the ordinary Nernst effect (ONE) of Bi$_{1-x}$Sb$_{x}$\cite{Roschewsky_ONE_R2w}, and the collective action of the spin Seebeck effect (SSE) in CoFeB followed by the inverse spin Hall effect (ISHE) in Bi$_{1-x}$Sb$_{x}$\cite{AvciPRB2014_R2w_thermoelectric}). The magnetic field amplitude dependence of $R_{2\omega}$ allows one to differentiate contributions from each effect: see MATERIALS AND METHODS for the details. $H_{\rm{DL}}$ ($H_{\rm{FL}}$) is related to the damping-like (field-like) spin Hall efficiency via $\xi_\textrm{DL(FL)}=\frac{2e}{\hbar}\frac{H_\textrm{DL(FL)}M_\textrm{s}t_\textrm{eff}}{j_\textrm{BiSb}}$, where $j_{\rm{BiSb}}$ is the current density in Bi$_{1-x}$Sb$_x$, $M_\textrm{s}$ and $t_\textrm{eff} \equiv t_\textrm{CoFeB} - t_\textrm{D}$ denote the saturation magnetization and the effective thickness of the CoFeB layer. $t_\textrm{D}$ is the thickness of the magnetic dead layer (see supplementary materials for details of magnetic properties of the films). From hereon, we discuss the values of $\xi_{\rm{DL}}$ and $\xi_{\rm{FL}}$.
%: see MATERIALS AND METHODS for the discussion on the thermoelectric effect contributions to $R_{2\omega}$. 
To estimate the spin Hall conductivity of Bi$_{1-x}$Sb$_{x}$, we use the relation $\sigma_\textrm{SH} = \xi_\textrm{DL} \sigma_\textrm{BiSb}$, where the spin transmission across the Bi$_{1-x}$Sb$_{x}$/CoFeB interface is assumed transparent\cite{Weiler2013PRL,RojasSanchez2014PRL,Zhang2015nphys}. Taking into account a non-transparent interface will result in larger $\xi_\textrm{DL}$ (and likely $\xi_\textrm{FL}$) and therefore results in larger $\sigma_\textrm{SH}$.

\subsection*{BiSb thickness dependence of $\sigma_\textrm{SH}$}
%Films of different \textit t$_\textrm{BiSb}$ and with a CoFeB wedge of \textit t$_\textrm{CoFeB}$ ranging from \num{0.5} to \SI{5}{\nano\meter} were grown using a moving linear shutter. These films were fabricated into Hall bar devices with
We first study the layer thickness dependence of the transport properties of heterostructures with nearly equiatomic Bi$_{0.53}$Sb$_{0.47}$ ($t_{\textrm{Bi}}$$\sim$ $t_{\textrm{Sb}}$$\sim$ 0.35 nm). The conductivity of Bi$_{0.53}$Sb$_{0.47}$ is plotted against $t_{\textrm{BiSb}}$ in Fig.~\ref{fig:Fig_2}(a). The slight increase of $\sigma_\textrm{BiSb}$ with $t_{\textrm{BiSb}}$ may be related to the larger grain size of thicker films that reduces the scattering at grain boundaries. Note that $\sigma_\textrm{CoFeB}$ takes an average value of $\sim$$\SI{5.5E3}{\per\ohm\per\centi\meter}$ and shows little dependence on $t_{\textrm{BiSb}}$. The $t_\textrm{BiSb}$ dependence of $\xi_\textrm{DL}$ and $\xi_\textrm{FL}$ for Bi$_{0.53}$Sb$_{0.47}$/CoFeB heterostructures measured at \SI{300}{\kelvin} are shown in Fig.~\ref{fig:Fig_2}(b) and \ref{fig:Fig_2}(d), respectively. For a given $t_\textrm{BiSb}$, we studied devices with three different $t_\textrm{CoFeB}$ ($\sim$\num{2}, \num{3.4}, and \SI{4.3}{\nano\meter}).
% (See MATERIALS AND METHODS for details of representative $R_{2\omega}$ results).
We find $\xi_\textrm{DL}$ of Bi$_{0.53}$Sb$_{0.47}$ is of the same sign with that of Pt \cite{HoffmannSHE_Review} and is consistent with previous reports on MBE-grown Bi$_{0.9}$Sb$_{0.1}$ \cite{NguyenNatMat2018_BiSb_MBE} and stoichiometric Bi$_2$Se$_3$ \cite{MellnikNature2014,Han2017PRL} topological insulators. At $t_\textrm{BiSb}$$\sim$$\SI{8}{\nano\meter}$, $\xi_\textrm{DL}$ reaches a maximum of $\sim$ 0.65. This value is significantly larger than those found in heavy metals but lower than some recent reports on Bi-based topological insulators \cite{MellnikNature2014,Wang2017,NguyenNatMat2018_BiSb_MBE,MahendraNatMat2018_BiSe_sputter_SOT}. $\xi_\textrm{DL}$ shows little dependence on $t_\textrm{CoFeB}$, indicating that the CoFeB layer plays little role in setting the SOT of the heterostructures.
%\textcolor{red}{(We don't show this relationship, it's difficult to understand the message here.)} \textcolor{blue}{(The sentence has been modified without referring to the equation)}
In order to take into account the change of $\sigma_\textrm{BiSb}$ with $t_\textrm{BiSb}$, we plot %apparent
%\textcolor{red}{(what's the meaning of "apparent"? we don't use this hereafter.)} \textcolor{blue}{(Because we didn't consider the interface transparency, this SHC is only an apparent value, not the intrinsic value of the material)}\textcolor{red}{(removed all "apparent" and "effective" $\sigma_\textrm{SH}$,  $\xi_\textrm{DL}$... since the definition is not explicitly described at the end of page 6.)}\textcolor{blue}{(For carrier mobility and density, effective is kept. The others have been removed.)}
the spin Hall conductivity $\sigma_\textrm{SH} = \xi_\textrm{DL} \sigma_\textrm{BiSb}$ against $t_\textrm{BiSb}$ in Fig.~\ref{fig:Fig_2}(c). $\sigma_\textrm{SH}$ increases with increasing $t_\textrm{BiSb}$ and tends to saturate beyond $t_\textrm{BiSb}$ of $\sim$8 nm. Such thickness dependence resembles that expected for the bulk SHE in Bi$_{0.53}$Sb$_{0.47}$ and is inconsistent with the surface-state-dominant scenario \cite{MellnikNature2014,ShiomiPRL2014} nor with the quantum confinement picture \cite{MahendraNatMat2018_BiSe_sputter_SOT}. We fit all data using the relation $\sigma_\textrm{SH} = \bar{\sigma}_\textrm{SH}[1-\textrm{sech}(\frac{t_\textrm{BiSb}}{\lambda})]$ with the \textit{bulk} spin Hall conductivity $\bar{\sigma}_\textrm{SH}$ and the spin diffusion length $\lambda$ as the fitting parameters\cite{Liu_2011_ST_FMR}. We find $\bar{\sigma}_\textrm{SH} = 496\pm26$ $(\hbar/e)\Omega^{-1}$cm$^{-1}$ and $\lambda = 2.3\pm0.4$ nm for Bi$_{0.53}$Sb$_{0.47}$.

%From Amp$\rm{\grave{e}}$re's Law, we estimate $H_{\rm{Oe}}/{j_{\rm{BiSb}}}\equiv2{\pi}t_{\rm{BiSb}} \times 10^{5}$ (in $\rm{Oe}/10^{6} A{\cdot}cm^{-2}$), which can then be subtracted from the total effective field to obtain $H_{\rm{FL}}$.
Figure~\ref{fig:Fig_2}(d) illustrates the $t_\textrm{BiSb}$ dependence of $\xi_\textrm{FL}$ for heterostructures with different CoFeB thicknesses. The contribution from $H_{\rm{Oe}}$, which takes the form of $H_{\rm{Oe}}/{j_{\rm{BiSb}}} = 2{\pi} t_{\rm{BiSb}}$ ($10^{-1} \ \rm{Oe} / (A \cdot cm^{-2}$))
%\textcolor{red}{(please check, I changed the formula)}\textcolor{blue}{($\cdot$ has been added; Units have been changed.)}
according to the Amp$\rm{\grave{e}}$re's law, is shown by the dotted lines in Fig.~\ref{fig:Fig_2}(d). Note that $H_{\rm{Oe}}$, being negative in our convention, is subtracted from the total field-like SOT to calculate $\xi_\textrm{FL}$. We find $H_{\rm{FL}}$ is opposite to the Oersted field and $\xi_{\rm{FL}}$ takes a constant value of $\sim$0.2 throughout the range of $t_\textrm{BiSb}$ studied. The sign of $\xi_\textrm{FL}$ for Bi$_{0.53}$Sb$_{0.47}$/CoFeB agrees with that of metallic Pt/Co/AlO$_x$\cite{GarelloNatNano2013_SOT}, while being opposite to that found in Bi$_2$Se$_3$/NiFe \cite{MellnikNature2014} and in MoS$_2$/CoFeB\cite{ShaoNanoletter2016}. The nearly constant $\xi_\textrm{FL}$ against $t_\textrm{BiSb}$ is observed for all structures with different $t_\textrm{CoFeB}$.
%As the maximum $t_\textrm{CoFeB}(= \SI{4.3}{\nano\meter})$ is larger than the transverse spin diffusion length of CoFeB ($\sim\SI{1}{\nano\meter}$)\cite{KimNatMat2012_Harmonic}, these results indicate that the CoFeB/MgO interface plays little role in defining $\xi_\textrm{FL}$ in this system, which is in contrast to the HM/CoFeB heterostructures\cite{KimNatMat2012_Harmonic,OuPRB2016_FL_SOT}.
The distinct $t_\textrm{BiSb}$ dependence of $\xi_\textrm{FL}$ and $\xi_\textrm{DL}$ suggests the two orthogonal components of SOT originate from phenomena of different characteristic length scales\cite{KimNatMat2012_Harmonic,OuPRB2016_FL_SOT}.

%$V_{\rm{const}}$ and $V_{\rm{ONE}}$ as a function of $I_{\rm{CoFeB}}^{2}$ and $I_{\rm{BiSb}}^{2}$ are plotted in Fig.~\ref{fig:Fig_2}(h) and (i), respectively. $I_{\rm{CoFeB}}$ and $I_{\rm{BiSb}}$ are calculated based on the parallel resistance model for Bi$_{0.53}$Sb$_{0.47}$/CoFeB bilayers of different thicknesses. We find that $V_{\rm{const}}$ ($V_{\rm{ONE}}$) is proportional to the power dissipation or the square of the current flowing within the CoFeB (Bi$_{0.53}$Sb$_{0.47}$) layer, reconfirming the thermoelectric origin of these contributions and the validity of our interpretation of $R_{2\omega}$.

\subsection*{Bi$_{1-x}$Sb$_x$ composition and facet dependence of $\sigma_\textrm{SH}$}

\begin{figure}
 \centering\includegraphics[width=0.85\columnwidth]{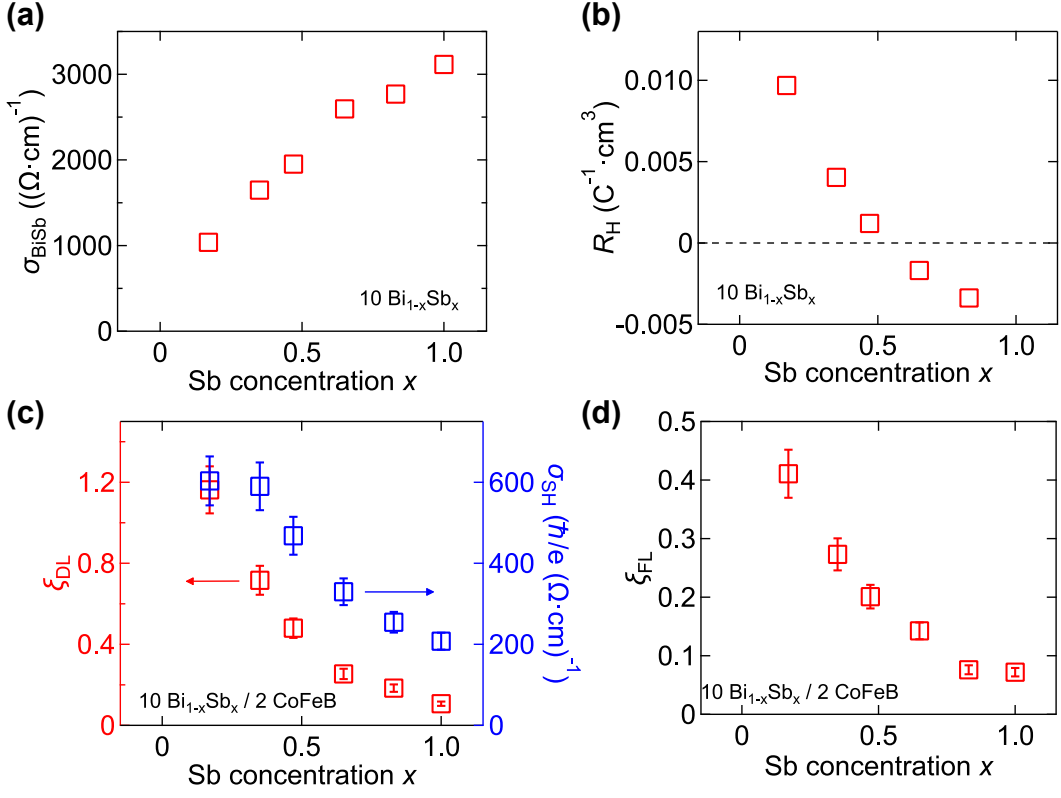}\\
  \caption{\textbf{Bi$_{1-x}$Sb$_x$ composition dependence of carrier transport and $\sigma_\textrm{SH}$.} (\textbf{a},\textbf{b}) Sb composition $x$ dependence of the conductivity $\sigma_\textrm{BiSb}$ (\textbf{a}) and the Hall coefficient $R_\textrm{H}$ (\textbf{b}) of Bi$_{1-x}$Sb$_x$ %\textcolor{brown}{(A665-06 for \bf{(a)} and \bf{(b)})}
  with $t_\textrm{BiSb}=10$ nm. For these studies, heterostructures without the CoFeB layer was used. (\textbf{c},\textbf{d}) Sb concentration ($x$) dependence of the damping-like spin Hall efficiency $\xi_\textrm{DL}$ (left axis), the spin Hall conductivity $\sigma_\textrm{SH}$ (right axis) (\textbf{c}), and the field-like spin Hall efficiency $\xi_\textrm{FL}$ (\textbf{d}) for 10 Bi$_{1-x}$Sb$_x$/2 CoFeB heterostructures. %\textcolor{brown}{(A665-07 for \bf{(c)} and \bf{(d)})}.
  All data were collected at \SI{300}{\kelvin}.% \textcolor{red}{Change x-axis to Sb concentration $x$.}
  %\textcolor{red}{Change y-axis unit expression for (a) as in Fig. 2c.)}\textcolor{blue}{(Done)}
  }
  \label{fig:Fig_3}
\end{figure}

Bulk Bi$_{1-x}$Sb$_x$ alloy is known to be a semiconductor with a small band gap hosting topological surface states
 for $0.09 \leq x \leq 0.22$ and is semimetallic for the other compositions\cite{Yim1972_BiSbbulk,Lenoir2001,Liu1995PRB,Teo2008PRB}. In an effort to shed light on the origin of the SOT, we investigate the Sb concentration ($x$) dependence of $\sigma_\textrm{SH}$ and related parameters in 10 Bi$_{1-x}$Sb$_{x}$/2 CoFeB heterostructures.
%Typical film structure consists of Sub./0.5 Ta/[0 - 0.6 Bi $\vert$ 0.6 - 0 Sb]$_{16}$/0.34 Bi/$t_\textrm{CoFeB}$ CoFeB/2 MgO/1 Ta. Bi$_{1-x}$Sb$_x$ alloy of varying $x$ with approximately a constant total thickness $t_\textrm{BiSb} \sim \SI{10}{\nano\meter}$ was grown by depositing alternate Bi and Sb wedges of opposite gradients.
To characterize the basic transport properties of Bi$_{1-x}$Sb$_x$ alloy, stacks without the CoFeB layer (\textit{i.e.} $t_\textrm{CoFeB}=\num{0}$) were also deposited and measured.
%AFM study suggests elemental Bi is forming practically discontinuous islands on Ta seed, leading to an anomalously high sheet resistance.
We have excluded pure Bi ($x=0$) from this study due to its large sheet resistance (considerably larger than those of the $x\neq 0$ alloys) and island-like morphology which may result in highly non-uniform current flow within the CoFeB layer. For alloys with $x>0$, the surface roughness significantly improves, as shown in Fig.~\ref{fig:Fig_1}(b).
%\textcolor{red}{(Throughout the paper, which data are from samples without the CoFeB layer? Can you specify in the figure captions for our purpose?.)}\textcolor{blue}{(Samples without CoFeB is used for determining the parameters for two-carrier model, i.e. the $x$ and $T$ dependence of conductivity, MR and $R_H$. The structures used for obtaining the data in the figures are indicated in the figures)}\textcolor{red}{What about the X-ray samples? Please write the exact structure together with the run number in the figure caption. We can delete them later.}\textcolor{blue}{(run number has been added in the figure captions emphasized in brown for all the composition dependent results)}
Figure~\ref{fig:Fig_3}(a) shows the $x$ dependence of $\sigma_\textrm{BiSb}$: $\sigma_\textrm{BiSb}$ increases monotonically with increasing Sb concentration.
%For $x = 0$ (\textit{i.e.} elementary Bi), the poor wetting of Bi on Ta-seeded SiO$_x$ surface resulting in almost discontinue island-like structure \cite{Hirose2018} which may explain the very low conductivity of the film. Upon adding a small amount of Sb in Bi, the wettability of the alloy significantly improves, leading to similar granular and continuous structures with that shown in the TEM in Fig.~\ref{fig:Fig_1} for $x\sim0.0.47$.
This is consistent with previous report on the transport properties of bulk Bi$_{1-x}$Sb$_x$ alloy\cite{Yim1972_BiSbbulk}, where it was shown that Bi$_{1-x}$Sb$_x$ gradually changes from being a semiconductor to a semimetal with increasing Sb concentration.
%For the following analysis, we will not consider Bi/CoFeB bilayer, having very different roughness and morphology compared with other samples.
As shown in Fig.~\ref{fig:Fig_3}(b), the ordinary Hall coefficient $R_\textrm{H} \equiv R_\textrm{xy}t_\textrm{BiSb}/H_\textrm{z}$ ($H_{\textrm{z}}$ is the external field $H_{\textrm{ext}}$ along $z$)
%\textcolor{red}{($H_{ext}$ or $H_z$?)}\textcolor{blue}{($H_z$)}
also varies monotonically with increasing $x$. In our convention, $R_\textrm{H} > 0$ $(R_\textrm{H}< 0)$ corresponds to carrier transport being dominated by electrons (holes). We find the carriers of Bi-rich alloys are electron-dominant whereas the Sb-rich structures are hole-dominant, accompanied by a smooth sign change of $R_\textrm{H}$ at $x$$\sim$0.55. This reflects the multi-carrier nature of the polycrystalline Bi$_{1-x}$Sb$_x$ films, having at least a hole and an electron pocket at the Fermi level. We note that this differs from the ternary (Bi$_{1-x}$Sb$_x$)$_2$Te$_3$ topological insulator for which $R_\textrm{H}$ diverges and abruptly changes sign when traversing the Dirac point\cite{KondouNatPhys2016}.

$\xi_\textrm{DL}$, $\sigma_\textrm{SH}$, and $\xi_\textrm{FL}$ as a function of $x$ for Bi$_{1-x}$Sb$_x$/2 CoFeB heterostructures are presented in Figs.~\ref{fig:Fig_3}(c) and \ref{fig:Fig_3}(d). $\xi_\textrm{DL}$ and $\xi_\textrm{FL}$ increase with increasing Bi concentration, reaching a maximum of $\xi_\textrm{DL}\sim 1.2$ and $\xi_\textrm{FL}\sim 0.41$ for structures with $x\sim0.17$,
a composition for which bulk Bi$_{1-x}$Sb$_x$ is commonly classified as a topological insulator. However, we emphasize that the BiSb thickness dependence in the previous section and the facet dependence of SHE in the next paragraph both suggest the bulk origin of the SHE.
The $x$ dependence of $\sigma_\textrm{SH}$ exhibits similar trend with that of $\xi_\textrm{DL}$: we find a plateau of $\sigma_\textrm{SH} \sim 600 (\hbar/e)\Omega^{-1}$cm$^{-1}$ for $x<0.35$. Interestingly, such $x$ dependence of $\sigma_\textrm{SH}$ is in very good agreement with that obtained from tight binding calculations\cite{SahinPRL2015_BiSb_SHC}, suggesting the dominance of the intrinsic contribution over that of the extrinsic skew scattering and side-jump contributions for the observed SHE in Bi$_{1-x}$Sb$_x$.

%\subsection*{Crystal orientation dependence of $R_{2\omega}$}
\begin{figure}
\centering
\includegraphics[width=0.85\columnwidth]{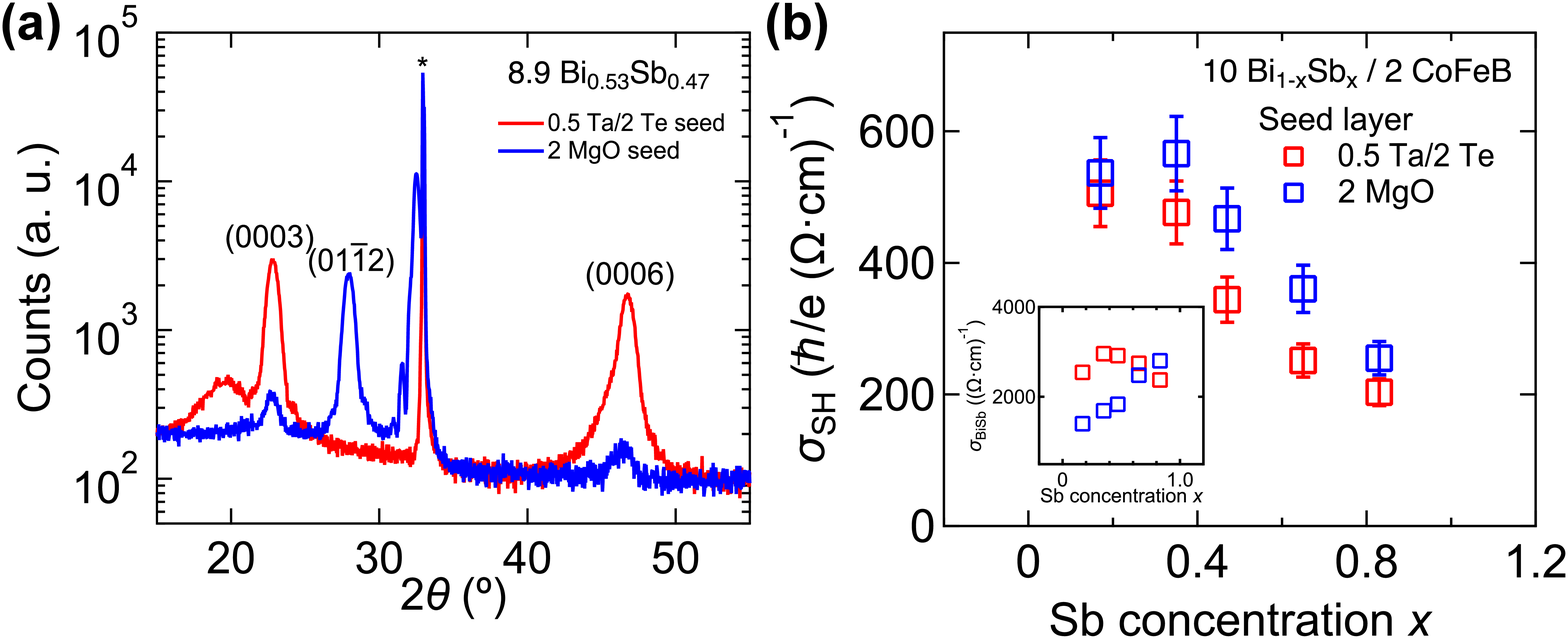}\\
\caption{\textbf{Facet dependence of $\sigma_\textrm{SH}$.} (\textbf{a}) X-ray diffraction (XRD) spectra for 8.9 Bi$_{0.53}$Sb$_{0.47}$ %\textcolor{brown}{(A629 series for \bf{(a)})}
grown on 0.5 Ta/2 Te (red) and 2 MgO (blue) seed layers. Heterostructures without the CoFeB layer was used. (\textbf{b}) Sb concentration ($x$) dependence of spin Hall conductivity $\sigma_\textrm{SH}$ for 10 Bi$_{1-x}$Sb$_x$/2 CoFeB heterostructures %\textcolor{brown}{(A665-07 for \bf{(b)})}
with the two seed layers described in (a). The inset shows the $x$ dependence of $\sigma_\textrm{BiSb}$.
%\textcolor{red}{(Show the x-axis of the inset. Reduce the number of ticks of the y-axis.)}\textcolor{blue}{(Done)}
The measurement temperature is \SI{300}{\kelvin}.
}
\label{fig:Fig_4}
\end{figure}

We have also varied the seed layer of the Bi$_{1-x}$Sb$_x$ layer to study the facet-dependent SHE.
%XRD of nearly equiatomic Bi$_{0.53}$Sb$_{0.47}$ films are used to characterize the texture of Bi$_{1-x}$Sb$_x$.
Figure~\ref{fig:Fig_4}(a) shows the XRD spectra of 8.9 nm thick Bi$_{0.53}$Sb$_{0.47}$ films grown on different seed layers, showing that the orientation of Bi$_{0.53}$Sb$_{0.47}$ nanocrystallites can be tuned from being practically random (Bi$_{0.53}$Sb$_{0.47}$ on 0.5 Ta seed; c.f. Fig.\ref{fig:Fig_1}(a)) to strongly (0003)-oriented (0.5 Ta/2 Te seed) or strongly (01$\bar1$2)-oriented (2 MgO seed). The Sb concentration ($x$) dependence of the longitudinal conductivity $\sigma_\textrm{BiSb}$
%\textcolor{red}{(unify to longitudinal conductivity elsewhere?)}\textcolor{blue}{(have defined in the first part.)}
and the SHC of Bi$_{1-x}$Sb$_x$ are shown in Fig.~\ref{fig:Fig_4}(b). As evident from the inset of Fig.~\ref{fig:Fig_4}(b), difference in the texture causes large changes in $\sigma_\textrm{BiSb}$, particularly at smaller $x$. Interestingly, however, the $x$ dependence of SHC (Fig.~\ref{fig:Fig_4}(b)) hardly changes upon varying the Bi$_{1-x}$Sb$_x$ texture and $\sigma_\textrm{BiSb}$. We thus infer that topological surface states, which are intimately related to the Bi$_{1-x}$Sb$_x$ facets \cite{TeoPRB2008}, are therefore unlikely to be the primary source of the observed SHE.
%\textcolor{red}{(Changes in the orientation may also result in changes in the defect concentration, which can influence the carrier density. So maybe this assumption is a bit less convincing. I'm not sure why the mobility independent SHC leads to bulk SHE.)} \textcolor{blue}{(Focus on intrinsic mechanism of the bulk band structure)}\textcolor{red}{(Removed the sentence "Assuming the modulation of $\sigma_\textrm{BiSb}$ is solely due to changes in the carrier mobility,")}
The robustness of SHC against $\sigma_\textrm{BiSb}$ further consolidates our suggestion that the intrinsic mechanism can account for the observed SHE.

\subsection*{Temperature dependence of $\sigma_\textrm{SH}$}
%conductivity $\sigma_\textrm{BiSb}$
\begin{figure*}
\centering
\includegraphics[width=1\columnwidth]{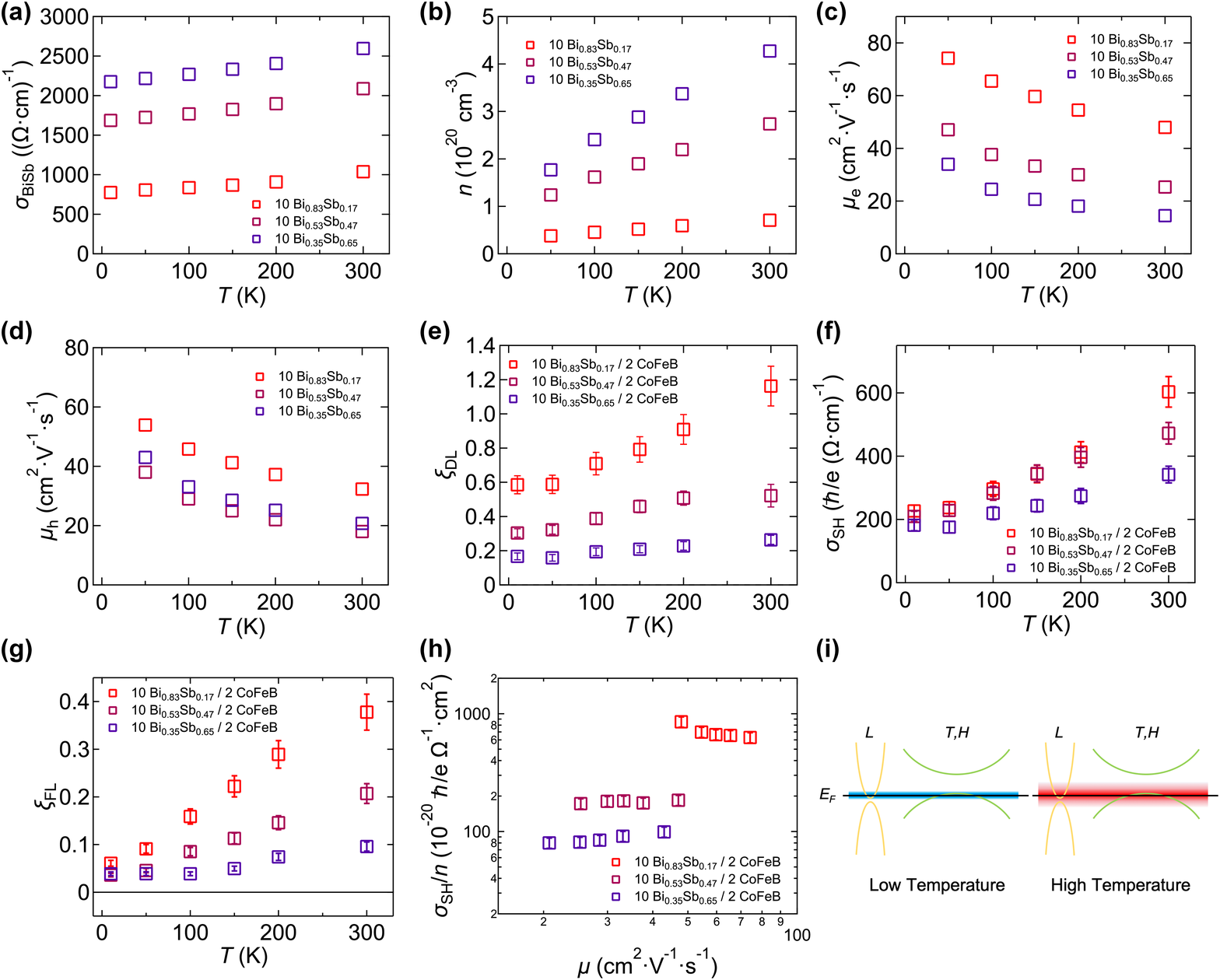}\\
\caption{\textbf{Temperature dependence of carrier transport and $\sigma_\textrm{SH}$.} (\textbf{a}-\textbf{c}) Temperature dependence of the conductivity $\sigma_{\rm{BiSb}}$ (\textbf{a}), the effective carrier concentration $n$ (\textbf{b}), the electron mobility $\mu_\textrm{e}$ (\textbf{c}), and the hole mobility $\mu_\textrm{d}$ (\textbf{c}) for 10 Bi$_{1-x}$Sb$_x$. Heterostructures without the CoFeB layer were used. %\textcolor{brown}{(A665-06 for \bf{(a)-(d)})}
(\textbf{e}-\textbf{g}) Temperature dependence of the damping-like spin Hall efficiency $\xi_\textrm{DL}$ (\textbf{e}), the spin Hall conductivity $\sigma_\textrm{SH}$ (\textbf{f}), and the field-like spin Hall efficiency $\xi_\textrm{FL}$ (\textbf{g}) for 10 Bi$_{1-x}$Sb$_x$/2 CoFeB heterostructures. %\textcolor{brown}{(A665-07 for \bf{(d)-(f)})}
 (\textbf{h}) $\sigma_{\rm{SH}}/n$ as a function of the majority carrier mobility ($\mu_\textrm{e}$ for $x$=0.17, 0.47 and $\mu_\textrm{h}$ for $x$=0.65) 10 Bi$_{1-x}$Sb$_x$/2 CoFeB heterostructures. (\textbf{i}) Schematic illustration of the band structures of Bi-rich Bi$_{1-x}$Sb$_x$ alloy with thermal broadening at low and high temperatures.}
\label{fig:Fig_5}
\end{figure*}

Finally, we examine the temperature dependence of the transport properties for Bi$_{1-x}$Sb$_x$/CoFeB heterostructures with selected $x$ ($x\sim$0.17, 0.47 and 0.65). Here the seed layer of Bi$_{1-x}$Sb$_x$ is 0.5 nm Ta. Figure~\ref{fig:Fig_5}(a) shows $\sigma_\textrm{BiSb}$ as a function of the measurement temperature. We find Bi$_{1-x}$Sb$_x$ possesses weak and positive temperature coefficient of the conductance, being typical for a semiconductor.
%Although the residual conductance differs by a factor of three, the residual conductance ratio $\sigma_\textrm{BiSb}(\SI{300}{\kelvin})/\sigma_\textrm{BiSb}(\SI{10}{\kelvin}) \sim 1.2$ is almost identical for the three compositions.
%\textcolor{red}{(does the rrr mean anything for semiconductors?)}\textcolor{blue}{(Is less trivial compared to metals because both $n$ and $\mu$ can induce changes in $\sigma$)}
To obtain the variation of the carrier concentration and mobility of Bi$_{1-x}$Sb$_x$, temperature dependence of the longitudinal magnetoresistance (MR) ratio ($(R_\textrm{xx}(H_\textrm{z})-R_\textrm{xx}(H_\textrm{z}=0))/R_\textrm{xx}(H_\textrm{z}=0)$) and $R_\textrm{xy}$ were measured against $H_\textrm{z}$. Within the framework of a two-band model\cite{Zhu2018JPCM}, we define $n_\textrm{h}$ ($n_\textrm{e}$) as the effective hole (electron) concentration of Bi$_{1-x}$Sb$_x$, with an effective mobility $\mu_\textrm{h}$ ($\mu_\textrm{e}$). Assuming equal population of the two carriers ($n_\textrm{h} = n_\textrm{e}$ = $n$)\cite{Lenoir2001}, we evaluate these parameters for temperatures ranging from \SIrange{50}{300}{\kelvin}, where the MR ratio and $R_\textrm{xy}$ are respectively quadratic and linear with $H_\textrm{z}$ up to \SI{8}{\tesla} (see supplementary materials for details of two-band model analysis). The temperature dependence of the carrier concentration and the mobility are summarized in Figs.~\ref{fig:Fig_5}(b-d). The carrier concentration increases with increasing temperature, which we infer is caused by the thermal broadening of the Fermi-Dirac distribution. %Assuming $n \propto \exp(-E_\textrm{g}/2k_\textrm{B}T)$ where $E_\textrm{g}$ is the effective band gap of Bi$_{1-x}$Sb$_x$, fits on data measured at temperatures ranging from \SI{150}{\kelvin} to \SI{300}{\kelvin}
%\textcolor{red}{(Why in this temperature range, why not 50-300 K? Should we show the fitting results?)}
%yield in average $E_\textrm{g} \sim \SI{20}{\milli\electronvolt}$ for the three samples
%\textcolor{red}{(show variation of Eg?)}\textcolor{blue}{(the fit is not very robust, perhaps we should delete this part?)}
%. This is comparable to the thermal fluctuation at \SI{300}{\kelvin} ($k_\textrm{B}T \approx \ \SI{26}{\milli\electronvolt}$), thus confirming the essential role of thermally excited electron-hole pairs in the enhancement of $\sigma_\textrm{BiSb}$.
In contrast, $\mu_\textrm{h}$ and $\mu_\textrm{e}$ both decrease with increasing temperature, obeying a power law that scales with $\sim T^{-0.5}$. These results indicate a competition between the impurity-mediated ($\propto T^{1.5}$) and electron-phonon scattering ($\propto T^{-1.5}$), with the latter being more dominant. Compared to the carrier concentration and mobility of the majority carrier for bulk single-crystal Bi ($n\sim \SI{4.6E17}{\per\centi\meter\cubed}$; $\mu_\textrm{e}\sim\SI{6E5}{\centi\meter\squared\per\volt\per\second}$)\cite{Michenaud1972} and Sb ($n\sim \SI{3.9E19}{\per\centi\meter\cubed}$; $\mu_\textrm{h}\sim\SI{2E4}{\centi\meter\squared\per\volt\per\second}$)\cite{Oktu1967} at \SI{77}{\kelvin}, $n$ of the Bi$_{1-x}$Sb$_x$ films studied here are one to two orders of magnitude higher while $\mu$ are orders of magnitude lower. These are expected for sputtered polycrystalline thin films that contain significantly higher defect density compared to that of the bulk samples\cite{EmotoPRB2016}.

The temperature dependence of $\xi_\textrm{DL}$, $\sigma_\textrm{SH}$ and $\xi_\textrm{FL}$ for 10 Bi$_{1-x}$Sb$_x$/2 CoFeB heterostructures are plotted in Figs.~\ref{fig:Fig_5}(e), (f) and (g) respectively. Surprisingly, all these quantities strongly enhance upon increasing the temperature from \SI{10}{\kelvin} to room temperature (\SI{300}{\kelvin}), suggesting that such enhancement is a rather generic feature for Bi$_{1-x}$Sb$_x$ alloy. Notably for Bi$_{0.83}$Sb$_{0.17}$, up to a three-fold (two-fold) enhancement is observed for $\sigma_\textrm{SH}$ ($\xi_\textrm{DL}$) over the investigated temperature interval. While similar increase of $\xi_\textrm{FL}$ with increasing temperature was previously reported in HM/FM heterostructures\cite{KimPRB2014_Tdep_Ta-CoFeB_SOT,OuPRB2016_FL_SOT}, such strong enhancement of technologically-important $\xi_\textrm{DL}$ and $\sigma_\textrm{SH}$ with increasing temperature has not been observed in metallic systems. We have also studied the temperature dependence of $\sigma_\textrm{SH}$ for thinner Bi$_{1-x}$Sb$_x$ films (5.6 Bi$_{0.53}$Sb$_{0.47}$/2 CoFeB).
%\textcolor{red}{(CoFeB thickness correct? Is this in the supps? No need to revise the supps, but for our purpose, can you show the plots in a separete ppt?)}\textcolor{blue}{(The data is not shown in supps. A ppt with this figures has been attached, showing similar property of 10 Bi$_{1-x}$Sb$_x$/2 CoFeB.)}.
We find similar temperature dependence of $\sigma_\textrm{SH}$ compared to that shown in Fig.~\ref{fig:Fig_5}(e), which suggests that the temperature dependence of $\xi_\textrm{DL}$, $\sigma_\textrm{SH}$ and $\xi_\textrm{FL}$ is not due to a temperature dependent spin diffusion length of Bi$_{1-x}$Sb$_x$. We have also verified that the CoFeB magnetization hardly changes within this temperature range (see supplementary materials),
%\textcolor{red}{(Is this shown in the supps?)}\textcolor{blue}{(The magnetic loop data for 2 CoFeB over Bi$_{53}$Bi$_{0.47}$ measured by SQUID at 300 K and 10 K has been shown in Fig.S2)}
reassuring that the change of $\xi_\textrm{DL}$ and $\xi_\textrm{FL}$ with temperature is caused by the modulation of the injected spin current.

\section*{DISCUSSION}
%\begin{figure*}
%\centering
%\includegraphics[width=0.85\columnwidth]{Fig_6.eps}\\
%\caption{\textbf{Scaling of $\sigma_\textrm{SH}$ with $\mu$ and $n$.} Spin Hall conductivity $\sigma_\textrm{SH}$ against mobility of majority carrier $\mu$ (\textbf{a}) and carrier concentration $n$ in log-log scales. Dotted lines are linear fits to the data.}
%\label{fig:SHC_plot}
%\end{figure*}
Within the Drude model, the longitudinal conductivity $\sigma_\textrm{BiSb}$ is proportional to the carrier concentration $n$ and the mobility $\mu$. $\mu$ is proportional to the relaxation time $\tau$ since $\mu=e\tau/m^*$ where $m^*$ is the effective mass. On varying the temperature, contribution from $n$ surpasses that of $\mu$ in Bi$_{1-x}$Sb$_x$, resulting in a positive temperature coefficient of the conductance, as shown in Fig.~\ref{fig:Fig_5}(a) .
For the SHC, by definition, the relaxation time dependence of $\sigma_\textrm{SH}$ provides a measure of the mechanism of the SHE: $\sigma_\textrm{SH} \sim \tau^1$ for the extrinsic skew scattering and $\sigma_\textrm{SH} \sim \tau^0$ when the intrinsic or side-jump mechanism dominates\cite{SinovaRMP_SHE_review}.
%[Commonly, the scaling relationship between $\sigma_\textrm{SH}$ and the longitudinal conductivity (assumed to be proportional to $\tau$) is used for revealing the underlying mechanism of the SHE. While this is a good approximation for metals, this approach is inappropriate for semiconductors/semimetals for which $n$ can vary drastically. In order to highlight the essential role of $n$ in influencing $\sigma_\textrm{SH}$ of Bi-Sb binary alloys, we first plot...] \textcolor{red}{(Doesn't the bad regime have semiconductor?)}
With regard to the relation between $\sigma_\textrm{SH}$ and $n$, the intrinsic contribution should scale with $n$ if the analogy with the anomalous Hall conductivity applies\cite{LeeScience2004}. Calculations suggest that similar scaling between $\sigma_\textrm{SH}$ and $n$ holds for the extrinsic mechanisms\cite{TsePRL2006}. We may thus take the ratio $\sigma_\textrm{SH}/n$ to eliminate the effect of $n$ on the temperature dependence of $\sigma_\textrm{SH}$: $\sigma_\textrm{SH}/n$ must be proportional to $\mu^{1}$ for the extrinsic skew scattering mechanism and is a constant for the intrinsic/side-jump mechanism. Figure~\ref{fig:Fig_5}(h) shows $\sigma_\textrm{SH}/n$ as a function of the mobility $\mu$ (here the mobility of the majority carrier is used, \textit{i.e.}, $\mu_\textrm{e}$ for $x$=0.17, 0.47 and $\mu_\textrm{h}$ for $x$=0.65). We find a relatively weak mobility dependence of $\sigma_\textrm{SH}/n$ for all alloy compositions studied. The slope of  $\sigma_\textrm{SH}/n$ vs. $\mu$ tends to increase as the Sb concentration increases.
%The $n$ dependence of $\sigma_\textrm{SH}$ also provides clues on the origin of SHE. In Fig.~\ref{fig:SHC_plot}(b), we plot $\sigma_\textrm{SH}$ against $n$ in a log-log scale. %The unusual augmentation of $\sigma_\textrm{SH}$ with temperature in Bi$_{1-x}$Sb$_x$ appears to be correlated with the increase of $n$ driven by the thermal broadening of the Fermi-Dirac distribution.
%In order to test the relationship $\sigma_\textrm{SH} \propto n^\gamma$, we perform linear fits and yield $\gamma$ that decreases with increasing $x$, from $\gamma\sim1.5$ for $x=0.17$ to $\gamma\sim0.7$ for $x=0.65$.
%A linear function of the form $\sigma_\textrm{SH} \propto n^\gamma$ is fitted to the data to yield the exponent $\gamma$. We find that $\gamma$ decreases with increasing $x$, from $\gamma\sim1.5$ for $x=0.17$ to $\gamma\sim0.7$ for $x=0.65$.
%The monotonic reduction of $\gamma$ and the slope of $\xi_\textrm{DL}$ vs. temperature as the Sb concentration is increased
These results indicate that the extrinsic skew scattering contribution is relatively weak for Bi-rich alloys with large intrinsic SHE but this contribution becomes non-negligible (but still smaller than the intrinsic one) with increasing $x$ (and $\sigma_\textrm{BiSb}$). Although this is reminiscent of the crossover from intrinsic to extrinsic SHE for metallic Pt upon tuning the resistivity of the metal\cite{SagastaPRB2016_Scaling_SHA_Pt}, we note that $\sigma_\textrm{BiSb}$ at the crossover is one to two orders of magnitude lower than that of Pt. Alternatively, we consider this crossover is a consequence of the band structure modification induced by Sb doping. As shown in the schematic band structure Fig.\ref{fig:Fig_5}(i), the transport in Bi-rich Bi$_{1-x}$Sb$_x$ alloy is dominated by the Dirac-like electrons at the $L$-point in the momentum space\cite{Yim1972_BiSbbulk,Lenoir2001,Liu1995PRB,Teo2008PRB,Fuseya2012jpsj}. Upon substituting Bi with Sb, holes from the $T$ and $H$ points with quadratic-like dispersion become increasingly important for the conduction, as shown by the $x$-dependence of $R_\textrm{H}$ in Fig.~\ref{fig:Fig_3}(b). Our experimental results indicate that Dirac-like $L$-electrons, in contrast to holes in $T$ and $H$-pockets with quadratic dispersion, is the key for achieving large intrinsic SHE in Bi$_{1-x}$Sb$_x$. We thus consider the large enhancement of SHC with temperature is caused by the increased number of $L$-electrons due to thermal broadening of the Fermi-Dirac distribution.

%We highlight that $\sigma_\textrm{SH}/n$ represents the average contribution of free charge carriers at the Fermi level for generating SHC.
Referring to the relation between the carrier density, mobility and conductivity that derives from the Drude model, $\sigma_\textrm{SH}/n$ can be regarded as the equivalent carrier mobility of transverse spin current.
%If we compare Bi$_{0.83}$Sb$_{0.17}$ with Bi$_{35}$Sb$_{65}$, the spin Hall conductivity $\sigma_\textrm{SH}$ differs by a factor of two, however, the equivalent mobility $\sigma_\textrm{SH}/n$ is larger for the former by nearly one order of magnitude.
To provide reference of the equivalent mobility, we estimate $\sigma_\textrm{SH}/n$ of a typical transition metal, Pt, which has the highest intrinsic spin Hall conductivity reported thus far ($\sigma_\textrm{SH} \approx 2000 \ (\hbar/e)\Omega^{-1}$cm$^{-1}$ at 0 K\cite{guo2008prl}). Assuming the carrier density $n$ of Pt is of the order of $10^{22} \ \textrm{cm}^{-3}$, we obtain $\sigma_\textrm{SH}/n \sim 20 \times 10^{-20} \ (\hbar/e)\Omega^{-1}$cm$^{2}$. This is more than an order of magnitude smaller than that of Bi$_{0.83}$Sb$_{0.17}$ evaluated at room temperature ($\sigma_\textrm{SH}/n \sim 800 \times 10^{-20} \ (\hbar/e)\Omega^{-1}$cm$^{2}$). The difference is also significant within the Bi$_{1-x}$Sb$_{x}$ alloy. If we compare Bi$_{0.83}$Sb$_{0.17}$ with Bi$_{0.35}$Sb$_{0.65}$, although the spin Hall conductivity $\sigma_\textrm{SH}$ differs by a factor of two, the equivalent mobility $\sigma_\textrm{SH}/n$ is larger for the former by nearly one order of magnitude.
These results demonstrate the exceptionally high spin current generation efficiency and mobility of the Dirac-like $L$-electrons in Bi-rich BiSb alloys compared to the majority holes in Sb-rich BiSb and the predominantly $s$-like conduction electrons in Pt.

In summary, we have studied the spin orbit torque (SOT) in sputter-deposited Bi$_{1-x}$Sb$_x$/CoFeB heterostructures.
The spin Hall conductivity (SHC) of Bi$_{1-x}$Sb$_x$ increases with increasing thickness until saturation and is facet independent.
These results suggest a dominant contribution from the bulk of the alloy: the effect of the topological surface states, if any, is not evident.
The SHC is the largest with Bi-rich composition and decreases with increasing Sb concentration.
Such trend is in accordance with the intrinsic spin Hall effect of Bi$_{1-x}$Sb$_x$ predicted using tight binding calculations.
Interestingly, the SHC and the damping-like spin Hall efficiency increase with increasing temperature. For example,  the damping-like spin Hall efficiency of Bi$_{0.83}$Sb$_{0.17}$ exhibits a two-fold enhancement from 5 K to room temperature, reaching $\xi_\textrm{DL} \sim 1.2$.
We infer that  thermally-excited population of the Dirac-like electrons in the $L$ valley of the narrow gap Bi$_{1-x}$Sb$_x$ is responsible for the temperature dependent SOT.
These results show that the Dirac-like electrons in Bi-rich Bi$_{1-x}$Sb$_x$ alloys are extremely effective in generating spin current and their equivalent spin current mobility is more than an order of magnitude larger than that of typical transition metals with strong spin orbit coupling.
%These electrons possess extremely high efficiency of spin current generation and large equivalent mobility.}
%The bulk intrinsic spin Hall effect can account for the magnitude, the thickness dependence, the Sb compositional dependence and facet dependence of SHE in Bi$_{1-x}$Sb$_x$. The thermal broadening of the Fermi-Dirac distribution function increases the carrier concentration of Dirac-like electrons at the $L$ point in the momentum space, leading to simultaneous increase of the longitudinal conductivity and the spin Hall conductivity.
%Unlike the longitudinal conductivity, the intrinsic spin Hall conductivity is unaffected by the reduction of the mobility at high temperature, resulting in an increase of the $\xi_\textrm{DL}$ ratio.
The very high spin Hall efficiency of Bi-rich Bi$_{1-x}$Sb$_x$ makes this material an outstanding candidate for applications involving spin current generation and detection at elevated temperatures. In addition, the lower carrier concentration and therefore a smaller electric-field screening length in Bi$_{1-x}$Sb$_x$ compared to the common heavy metals allow efficient electric field control of SOT, thus paving a route to multifunctional spinorbitronic devices being sensitive to external stimuli such as heat and electric field.

%Methods
\section*{MATERIALS AND METHODS}
\subsection*{Sample preparation and characterization}
All samples were grown at ambient temperature by magnetron sputtering on Si substrates (10 $\times$ 10 mm$^2$) coated with 100 nm thick thermally oxidized Si layer.
%The base pressure of of the system is $2\times10^{-7}$ Pa.
%The Ta seed layer were sputtered with a 50 W r.f. power under a 0.5 Pa Ar pressure. Bi and Sb were sputtered with rf power of 12 W and 15 W, respectively. The Ar gas pressure was set to $\sim$1 Pa during the sputtering. The CoFeB, Ta and MgO layers were sputtered at Ar gas pressure of 0.5 Pa using 50 W and 100 W rf power, respectively.
%\subsection*{Samples characterization and Fabrication of devices}
Atomic force microscopy (AFM) were used to characterize the roughness of the surface. $\theta - 2\theta$ X-ray diffraction (XRD) spectra were obtained using a Cu $K\alpha$ source in parallel beam configuration and with a graphite monochromator on the detector side. The saturation magnetization and the magnetic dead layer thickness of the CoFeB layer in the heterostructures were determined by hysteresis loop measurements using vibrating sample magnetometer (VSM). AFM, XRD and VSM studies were performed using unpatterned constant thickness films. High-angle annular dark-field scanning transmission electron microscopy (HAADF-STEM) analysis of cross-sectioned samples were performed using a FEI Titan G2 80-200 transmission electron microscope (TEM) with a probe forming spherical aberration corrector operated at 200 kV. The samples were cross-sectioned from a plain film into thin lamellae by focused ion beam (FIB) lift-out technique using FEI Helios G4 UX.
Hall bars for the transport measurements were pattered from the films using optical lithography and Ar ion etching. The width $w$ and the distance between the two longitudinal voltage probes $L$ are 10 $\mu$m and 25 $\mu$m, respectively. Contact pads to the Hall bars, 10 Ta /100 Au (thickness in nm), were formed using a standard lift-off processes.

\subsection*{SOT measurements}
We treat the CoFeB magnetization as a spin domain magnet with a magnetization vector $\bm{M}$ lying in the film plane ($xy$ plane) at equilibrium. The external magnetic field $\bm{H}_{\rm{ext}}$ is applied along the film plane with an angle $\varphi_H$ with respect to the $x$ axis. We assume the in-plane magnetic anisotropy of the CoFeB layer is negligible compared to the magnitude of $\bm{H}_{\rm{ext}}$. Thus the angle $\varphi$ between $\bm{M}$ and the $x$ axis is assumed be equal to $\varphi_H$.
%The electrical current flowing along $\bm{x}$ is converted to spin current via the spin Hall effect with the electron direction defined with $\boldsymbol{\sigma}$.  pointing along $\bm{y}$ at the Bi$_{1-x}$Sb$_{x}$/CoFeB interface via the spin Hall effect of Bi$_{1-x}$Sb$_{x}$.

When current is passed along $\bm{x}$, electrons with their spin direction parallel to $\bm{y}$ diffuses into the CoFeB layer via the spin Hall effect. The impinging spin current exerts spin-transfer torque (or often referred to as the spin-orbit torque) on the CoFeB magnetization. The torque can be decomposed into two components, the damping-like and field-like torques: the equivalent effective fields are defined as $H_{\rm{DL}}$ and $H_{\rm{FL}}$, respectively. Together with the Oersted field $H_{\rm{Oe}}$, $H_{\rm{DL}}$ and $H_{\rm{FL}}$ cause tilting of the CoFeB layer magnetization. When an ac current (amplitude $I_0$, frequency $\omega / 2 \pi$) is applied to the heterostructure, current-induced oscillation of $\bm{M}$ leads to Hall voltage oscillation via anomalous Hall effect (AHE) and planar Hall effect (PHE). The first harmonic (fundamental) voltage $V_{1\omega}$ represents the magnetization direction at equilibrium and the out-of-phase second harmonic voltage $V_{2\omega}$ provides information on the current-induced effective fields acting on the magnetization. We define $R_{i\omega} \equiv V_{i\omega} / (I_0/\sqrt{2})$ ($i=1,2$) to represent the harmonic signals.

Contributions to $R_{2\omega}$ include five terms. $R_{\rm{DL}}$, $R_{\rm{FL}}$, $R_{\rm{Oe}}$, which reflect changes in $R_{2\omega}$ caused by $H_{\rm{DL}}$, $H_{\rm{FL}}$ and $H_{\rm{Oe}}$, respectively, decay with increasing $H_{\rm{ext}}$. $R_{\rm{DL}}$ is proportional to the $x$ component of the magnetization ($\cos\varphi$) whereas $R_{\rm{FL}}+R_{\rm{Oe}}$ scales with $\cos2\varphi \cos\varphi$ due to the combined influences of AHE and PHE\cite{Kawaguchi_2013,AvciPRB2014_R2w_thermoelectric}. Current induced Joule heating and the different thermal conductivity of the substrate and air can lead to an out-of-plane temperature gradient\cite{AvciPRB2014_R2w_thermoelectric} across the heterostructure. With the temperature gradient, the anomalous Nernst effect (ANE) of CoFeB and the collective action of the spin Seebeck effect (SSE)\cite{Uchida2008nature} in CoFeB followed by the inverse spin Hall effect (ISHE) in Bi$_{1-x}$Sb$_{x}$ result in a contribution ($R_{\rm{const}}$) that does not depend on the size of $H_{\rm{ext}}$. Applying a field orthogonal to the out-of-plane temperature gradient produces the last term, $R_{\rm{ONE}}$, due to the ordinary Nernst effect (ONE)\cite{Roschewsky_ONE_R2w}. $R_{\rm{ONE}}$ scales linearly with $H_{\rm{ext}}$. The ONE of CoFeB is negligible compared to that of Bi$_{1-x}$Sb$_{x}$ due to the difference in the carrier density. Both $R_{\rm{const}}$ and $R_{\rm{ONE}}$ scale with $\cos\varphi$.

Putting together these contributions (and assuming $\varphi\sim\varphi_H$), $R_{2\omega}$ reads:
\begin{equation}
\begin{split}
\label{eq:R2w}
R_{2\omega}=&R_{\rm{DL}}+R_{\rm{ONE}}+R_{\rm{const}}+R_{\rm{FL}}+R_{\rm{Oe}}\\
=&\left(R_{\rm{AHE}}\dfrac{H_{DL}}{H_{\rm{ext}}+H_{\rm{K}}}+\dfrac{\mathcal{N}w{\Delta}T}{I_{0}}H_{\rm{ext}}+\dfrac{{\alpha}w{\Delta}T}{I_{0}}\right)\cos\varphi_H\\
& \ \ -2R_{\rm{PHE}}\dfrac{H_{\rm{FL}}+H_{\rm{Oe}}}{H_{\rm{ext}}}\cos2\varphi_H\cos\varphi_H,
\end{split}
\end{equation}
where $R_{\rm{AHE}}$ is the anomalous Hall resistance, $R_{\rm{PHE}}$ is the planar Hall resistance, $H_{\rm{K}}$ is the out-of-plane anisotropy field, $\mathcal{N}$ is the ONE coefficient of Bi$_{1-x}$Sb$_{x}$ and $\alpha$ is a coefficient that reflects the size of ANE and the combined action of SSE and ISHE. The distinct $H_{\rm{ext}}$ and $\varphi_H$ dependence of $R_{2\omega}$ for these contributions allows unambiguous separation of each term from the raw $R_{2\omega}$ signal. We first decompose $R_{2\omega}$ into two contributions of different $\varphi_H$ dependence, \textit{i.e.}  $\cos\varphi_H$ and $\cos2\varphi_H\cos\varphi_H$, and define the prefactors of these two parts as $A$ and $B$, respectively:
\begin{equation}
\label{eq:A}
\centering
A \equiv R_{\rm{AHE}}\dfrac{H_{DL}}{H_{\rm{ext}}+H_{\rm{K}}}+\dfrac{V_{\rm{ONE}}}{I_{0}}H_{\rm{ext}}+\dfrac{V_{\rm{const}}}{I_{0}},
\end{equation}
\begin{equation}
\label{eq:B}
B \equiv -2R_{\rm{PHE}}\dfrac{H_{\rm{FL}}+H_{\rm{Oe}}}{H_{\rm{ext}}}.
\end{equation}
Two parameters $V_{\rm{ONE}} \equiv Nw{\Delta}T$ and $V_{\rm{const}} \equiv {\alpha}w{\Delta}T$ are defined to describe the thermoelectric contributions.
The $H_{\rm{ext}}$ dependence of $A$ and $B$ are then fitted based on Eqs.~(\ref{eq:A}) and (\ref{eq:B}), respectively.
%We emphasize that the thermoelectric contributions described here interfere with anomalous-Hall-resistance-based metrology for quantifying SOT, including but not limited to the harmonic Hall and the d.c. differential Hall techniques \cite{LiuPRL2012_Pt_switching,Kawaguchi_2013}.

%\subsection*{Transport measurements}
\begin{figure*}
\centering
\includegraphics[width=1.0\columnwidth]{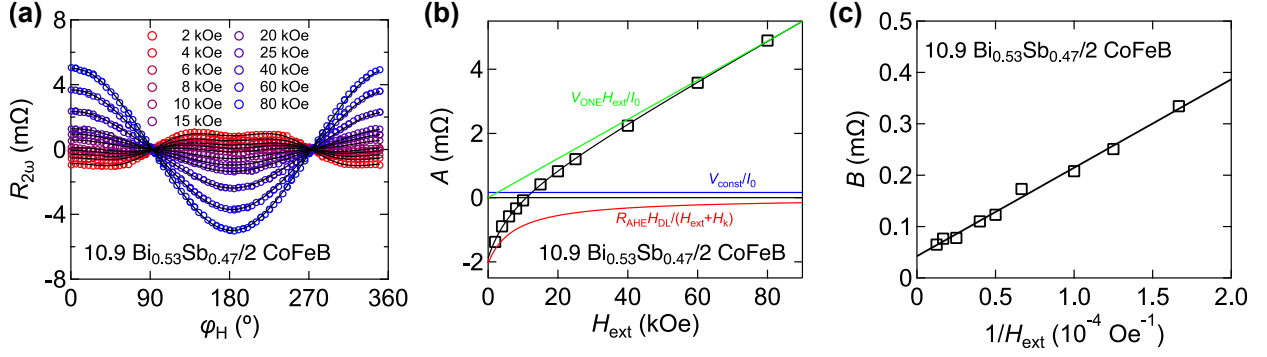}\\
\caption{\textbf{Representative second harmonic Hall resistances.} (\textbf{a}) Field angle $\varphi_H$-dependence of the second harmonic Hall resistance $R_{2\omega}$ for 10.9 Bi$_{0.53}$Sb$_{0.47}$/2 CoFeB heterostructure obtained using different $H_\textrm{ext}$ measured at \SI{300}{\kelvin}. (\textbf{b}) $H_\textrm{ext}$ dependence of the fitting parameter $A$. (\textbf{c}) $1/H_\textrm{ext}$ dependence of the fitting parameter $B$. In (b) and (c), the colored lines show contributions from each component described in Eqs.~(\ref{eq:A}) and (\ref{eq:B}): the black solid line show the sum of all contributions.}
\label{fig:Fig_6}
\end{figure*}

%A physical property measurement system (PPMS) was used to perform the dc transport measurement. A 20 $\mu$A d.c. current was applied to measure the anomalous Hall resistance. For the second harmonic Hall resistance measurement, a Keithley 6221 Sourcemeter was used as the a.c. source. A 17.5 Hz frequency and 1.5 mA r.m.s. amplitude a.c. current was supplied by the source. Two Stanford research systems (SRS) SR830 lock-in amplifiers were used to measure the first and second harmonic Hall voltages ($V_{\omega}$, $V_{2\omega}$) separately. We define $R_{\omega}{\equiv}V_{\omega}/I_{0}$ and $R_{2\omega}{\equiv}V_{2\omega}/I_{0}$. The harmonic Hall resistances were gotten by dividing the Hall voltages with the r.m.s. value of a.c. current.  The PPMS was also used to supply the high magnetic field in a.c. measurements.

Representative $R_{2\omega}$ as a function of $\varphi_H$ obtained using different $H_{\rm{ext}}$ for 10.9 Bi$_{0.53}$Sb$_{0.47}$/2 CoFeB heterostructure is shown in Fig.~\ref{fig:Fig_6}(a). Solid lines in the figure are fits to the data using Eq.~(\ref{eq:R2w}). See supplementary materials for the $\varphi_H$ dependence of $R_{1\omega}$. All data shown in this paper are collected using $\omega/2\pi=17.5$ Hz. The current amplitude is typically set to $\sim$1.5 mA$_{\textrm{rms}}$, which corresponds to a current density in the Bi$_{1-x}$Sb$_{x}$ layer of $\sim$$1\times10^{10}$ A/m$^2$. We find $R_{2\omega}$ scales linearly with current (data not shown).
%\textcolor{red}{(Is the current always fixed?)}\textcolor{blue}{(Yes, we used a AC source and fixed the current for all the samples and temperature.)}\textcolor{red}{Have you checked the current dependence? Is the current density more or less the same for all films (constant voltage will give constant J but constant current can cause significant changes in the current density)?}\textcolor{blue}{(We used a ammeter to monitor the current in the sample and the harmonic resistivity is obtained by harmonic voltage over the monitored current for every point. We have checked the current does not change too much in all the measurement.)}
%As $H_{\rm{ext}}$ varies from \num{2} to \SI{80}{\kilo\Oersted}, the size of $\varphi$-dependent $R_{2\omega}$ reduces and eventually $R_{2\omega}$ changes sign for sufficiently large $H_{\rm{ext}}$. The sign reversal of $R_{2\omega}$ cannot be explained by the SOT contribution alone, thus confirming the importance of including thermoelectric contributions in the analysis.
The $H_{\rm{ext}}$ dependence of one of the fitting parameters $A$ is shown in Fig.~\ref{fig:Fig_6}(b). The best fit to $A$ against $H_{\rm{ext}}$ is shown by the solid black line. The colored lines represent decomposition of each contribution following Eq.~(\ref{eq:A}) (see supplementary materials for determination of $R_{\rm{AHE}}$ and $H_{\rm{K}}$). In the small $H_{\rm{ext}}$ range, $R_{\rm{DL}}$ term (red line) dominates, whereas at larger field, $A$ changes sign and is eventually dominated by $R_{\rm{ONE}}$ (green line). $B$ is plotted against $1/H_{\rm{ext}}$ in Fig.~\ref{fig:Fig_6}(c) with the black solid line showing the best linear fit to the data using Eq.~(\ref{eq:B}).
%for $H_\textrm{ext}$ ranging from $\num{6}$ to $\SI{15}{\kilo\Oersted}$ according to
We extract $H_{\rm{DL}}$, $H_{\rm{FL}} + H_{\rm{Oe}}$, $V_{\rm{ONE}}$, and $V_{\rm{const}}$ from the two fits.
% and repeat the procedure for other thickness combinations of [Bi$\vert$Sb]/CoFeB bilayers.
We find $V_{\rm{const}}$ and $V_{\rm{ONE}}$ to be proportional to the square of the current flowing within the CoFeB and Bi$_{1-x}$Sb$_{x}$ layer, respectively. These results confirm the thermoelectric origin of these contributions and the validity of the interpretation of $R_{2\omega}$.\\
\\

\renewcommand{\baselinestretch}{2}
\renewcommand{\theequation}{S\arabic{equation}}
\renewcommand {\thefigure} {S\arabic{figure}}
\renewcommand {\thesection} {S\arabic{section}}
\makeatletter
\renewcommand\section{\@startsection {section}{1}{\z@}%
    {-1.5ex \@plus -1ex \@minus -.1ex}%
    {1ex \@plus.0005ex}%
    {\normalfont\large\bfseries}}
\makeatother
\setcounter{figure}{0}   

\section*{SUPPLEMENTARY MATERIALS}
Supplementary material for this article is available at...\\
S1. STEM results of films.\\
S2. Magnetic properties of BiSb/CoFeB.\\
S3. Anomalous Hall resistance and anisotropy field.\\
S4. First harmonic Hall resistance of BiSb/CoFeB.\\
S5. Two-band model analysis of BiSb.\\
S6. Evaluation of the SOT analysis protocol.\\
S7. The efficiency of BiSb SOT.\\
Figure S1. HAADF-STEM and EDS mapping.\\
Figure S2. Saturation magnetization and magnetic dead layer thickness.\\
Figure S3. Anomalous Hall resistance and anisotropy field.\\
Figure S4. First harmonic Hall resistance.\\
Figure S5. Temperature dependence of magneto-transport properties of Bi$_{1-x}$Sb$_x$.\\
Figure S6. SOT measurements of a standard sample: Pt/CoFeB.\\
References \cite{XuActa_2018,Ikeda_2010,Sinha2013apl,Kim2016PRL,Cho_2015,AliNature2014,Wang2014APL,lee2014apl,PaiAPL2012_SHE_W,garello2018ieee}.\\
%\cite{XuActa_2018,LauJJAP2017_Harmonic_IP,Ikeda_2010,Sinha2013apl,Kim2016PRL,Cho_2015,Chen2013PRB,AliNature2014,Wang2014APL,SagastaPRB2016_Scaling_SHA_Pt,lee2014apl,PaiAPL2012_SHE_W,garello2018ieee}.

%Acknowledgements
%\section*{}
%\noindent
\textbf{Acknowledgements}: We thank Y. Fuseya, H. Kohno, G. Qu for helpful discussions.
\textbf{Fundings}:  This work was partly supported by JSPS Grant-in-Aid for Scientific Research (Grant No. 16H03853), Specially Promoted Research (Grant No. 15H05702), Casio Science Foundation, and the Center for Spintronics Research Network (CSRN). Z.C. acknowledges financial support from Materials Education program for the future leaders in Research, Industry, and Technology (MERIT). Y.-C.L. is supported by JSPS International Fellowship for Research in Japan (Grant No. JP17F17064).
\textbf{Author Contributions}: Y.-C.L., Z.C. M.H. planned the study. Z.C. and Y.-C.L. grew the samples, designed the experimental set-up, performed electrical measurements and carried out data analysis. Y.-C.L. performed structural  characterization (AFM, XRD), fabricated Hall bar devices, and modeled the system transport properties. X.-D.X., T.O. and K.H. carried out TEM imaging. Z.C. and Y.-C.L. wrote the manuscript with input from M.H. All authors contributed to the discussion of results and commented on the manuscript.
\textbf{Competing interests}: The authors declare no competing interests.
\textbf{Data and materials availability}: All data needed to evaluate the conclusions in the paper are present in the paper and/or the supplementary materials. Additional data related to this paper may be requested from the corresponding authors.

% reference

%SUPPLEMENTARY MATERIAL *******************************************************************************************
\clearpage
\section{STEM \MakeLowercase{results of films}}
\label{sec:STEM}
We performed high-angle annular dark-field scanning transmission electron microscopy (HAADF-STEM) analysis of samples using a FEI Titan$^{\rm{TM}}$ G2 80-200 transmission electron microscope with a probe forming aberration corrector operated at 200 kV. Samples were cross-sectioned into thin lamellae by focused ion beam (FIB) lift-out technique using FEI Helios G4 UX. Nanobeam electron diffraction patterns were taken with a 10-$\rm{\mu}$m-diamter condenser aperture. Element-selected analysis were carried out via energy-dispersive X-ray spectroscopy (EDS)\cite{XuActa_2018}.

\begin{figure}
\begin{center}
\includegraphics[width=6in]{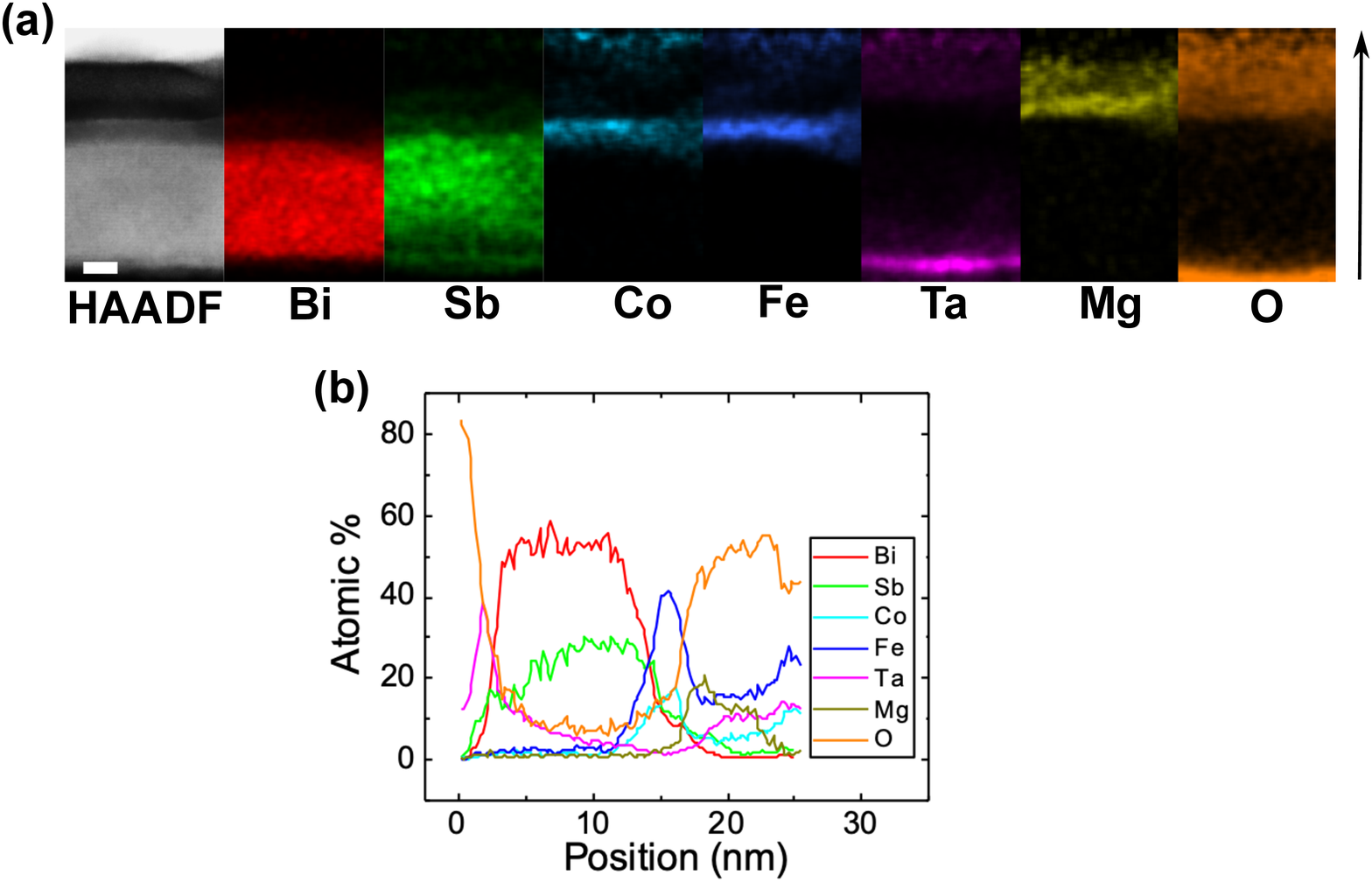}\\
\caption{\textbf{HAADF-STEM and EDS mapping.} (a) HAADF-STEM and elemental EDS maps of a heterostructure with 10.9 Bi$_{0.53}$Sb$_{0.47}$/2 CoFeB. The length of the horizontal white bar corresponds to $\sim$2 nm in the image. (b) Position dependence of the atomic ratio for different elements. The black arrow in (a) indicates the direction of the line scans. Position zero corresponds to the substrate/heterostructure interface. }
\label{fig:STEM}
\end{center}
\end{figure}

Figure~\ref{fig:STEM}(a) shows a HAADF-STEM image and the corresponding EDS maps of one of the heterostructures with 10.9 Bi$_{0.53}$Sb$_{0.47}$/2 CoFeB. EDS line scans averaged over the entire image (19 $\times$ 26 nm$^2$) are shown in Fig.~\ref{fig:STEM}(b). We find that Sb tends to diffuse toward the CoFeB layer. This may be attributed to the finite solubility of Sb in Co and Fe, in contrast to Bi which is practically immiscible with any of these elements.

\section{M\MakeLowercase{agnetic properties of} B\MakeLowercase{i}S\MakeLowercase{b}/C\MakeLowercase{o}F\MakeLowercase{e}B}
\label{sec:Mag}
\begin{figure}
\begin{center}
\includegraphics[width=6.5in]{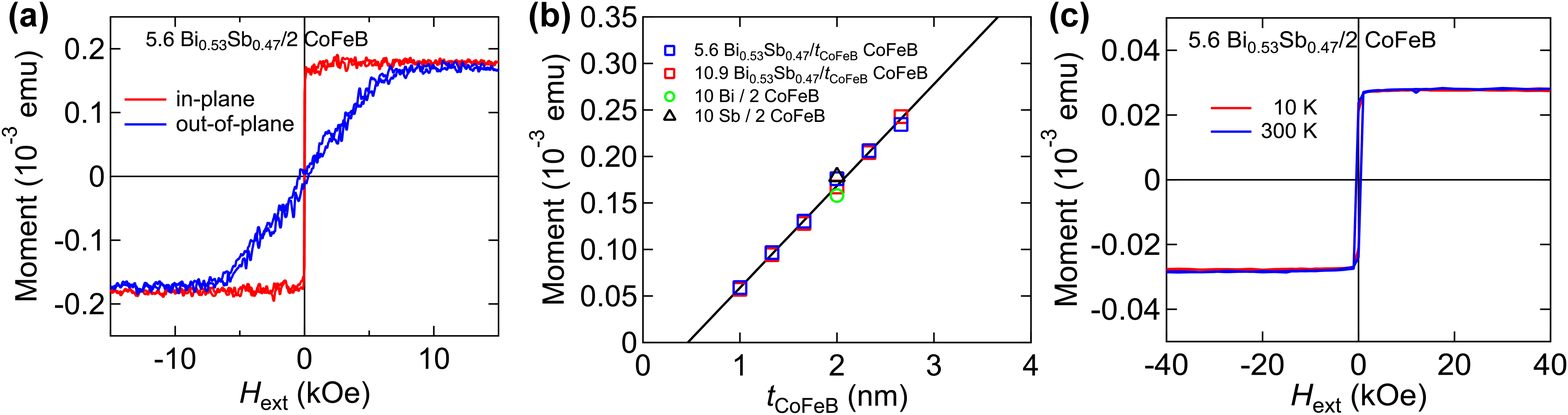}\\
\caption{\textbf{Saturation magnetization and magnetic dead layer thickness.} (a) Magnetization hysteresis loops of a heterostructure with 5.6 Bi$_{0.53}$Sb$_{0.47}$/2 CoFeB. The red and blue lines show the hysteresis when the field is applied along and normal to the film plane, respectively. (b) CoFeB thickness dependence of the saturated magnetic moment for heterostructures with 5.6 Bi$_{0.53}$Sb$_{0.47}$/$t_{\rm{CoFeB}}$ CoFeB (blue squares) and 10.9 Bi$_{0.53}$Sb$_{0.47}$/$t_{\rm{CoFeB}}$ CoFeB (red squares). The solid line represents linear fit to the data. The slope and the $x$-axis intercept of the linear line allows one to determine the saturation magnetization and the magnetic dead layer thickness, respectively. Data from heterostructures with 10 Bi/2 CoFeB and 10 Sb/2 CoFeB are laid together using green circle and black triangle, respectively. (c) In-plane magnetization hysteresis loops of a heterostructure with 5.6 Bi$_{0.53}$Sb$_{0.47}$/2 CoFeB. The red and blue lines represent hysteresis loops obtained at measurement temperatures of 10 K and 300 K, respectively. The results shown in (a) and (b) are obtained using VSM: the specimen film area is $\sim$9.6$\times$9.6 mm$^2$. SQUID is used to obtain the results shown in (c): the specimen film area is roughly 2$\times$6 mm$^2$.
%\textcolor{red}{(c) Use the same y-axis title as in (a).}\textcolor{blue}{(The sample is cut for SQUID measurements. We estimated the size of sample and obtain the magnetization. It should make no sense to use Moment here. For (a), we used Moment as y-axis because we need this value to determine the effective magnetization and magnetic dead layer in (b).)}
}
\label{fig:moment}
\end{center}
\end{figure}

%BiSb/CoFeB bilayers with different CoFeB thicknesses were prepared for characterizing the magnetic properties.
Magnetic moments of the heterostructures were measured using a vibrating sample magnetometer (VSM) at room temperature.  A superconducting quantum interference device (SQUID) is used to measure the temperature dependence of the magnetic properties. The magnetization hysteresis loops of a heterostructure with 5.6 Bi$_{0.53}$Sb$_{0.47}$/2 CoFeB are displayed in Fig.~\ref{fig:moment}(a) under an external magnetic field applied parallel and perpendicular to film plane. The magnetic easy axis of the CoFeB layer points along the film plane.

The $t_\textrm{CoFeB}$ dependence of the saturated magnetic moment of heterostructures with two different Bi$_{0.53}$Sb$_{0.47}$ layer thicknesses are plotted in Fig.~\ref{fig:moment}(b). Data from the two series are hardly discernable. We therefore assume that the saturation magnetization $M_{s}$ and the magnetic dead layer thickness $t_{\rm{D}}$ of the CoFeB layer are independent of the Bi$_{1-x}$Sb$_{x}$ thickness. Linear fit to all the data are shown in Fig.~\ref{fig:moment}(b) as the solid line. From the slope and the $x$-intercept of the linear line, $M_{s}$ and $t_{\rm{D}}$ are determined to be $1190\pm20{\;}\rm{emu/cm^{3}}$ and $0.46\pm0.04{\;}\rm{nm}$, respectively. In Fig.~\ref{fig:moment}(b), we also show $M_{s}$ values of heterostructures with 10 Bi/2 CoFeB and 10 Sb/2 CoFeB. As we find almost no difference in $M_{s}$ when the Sb concentration is changed, we assume a constant $M_{\rm{s}}$ and $t_{\rm{D}}$ for all heterostructures studied (with different $x$).

Figure~\ref{fig:moment}(c) shows the magnetization hysteresis loops of the same heterostructure shown in (a) evaluated at 300 K and 10 K. A SQUID magnetometer is used: the field is applied along the film plane. The two loops measured at 300 K and 10 K overlap. We thus assume the temperature has little influence on $M_{\rm{s}}$ of CoFeB for all samples.

\section{A\MakeLowercase{nomalous} H\MakeLowercase{all resistance and anisotropy field}}
\label{sec:DC_measure}
\begin{figure}
\begin{center}
\includegraphics[width=6.5in]{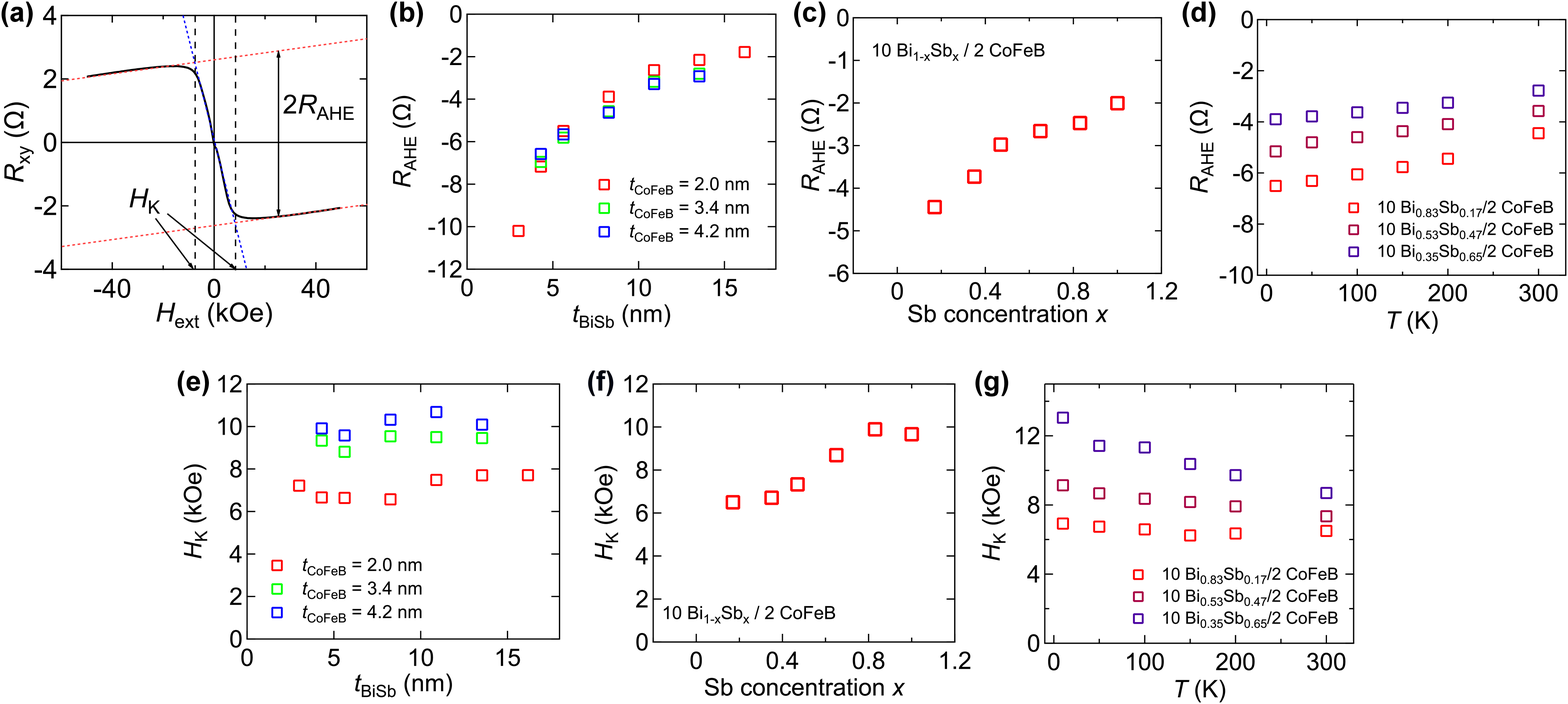}\\
\caption{\textbf{Anomalous Hall resistance and anisotropy field.} (a) Hall resistance $R_{\rm{xy}}$ of a heterostructure with 10.9 Bi$_{0.53}$Sb$_{0.47}$/2 CoFeB plotted as a function of magnetic field $H_\textrm{ext}$ along the film normal (along the $z$-axis). Red and blue dashed lines show linear fits to the data in the high-field and low-field ranges, respectively. Definitions of the anomalous Hall resistance$R_{\rm{AHE}}$ and the anisotropy field $H_{\rm{K}}$ are illustrated in the panel. (b-g) $R_{\rm{AHE}}$ and $H_{\rm{K}}$ of heterostructures with $t_\textrm{BiSb}$ Bi$_{1-x}$Sb$_{x}$/$t_\textrm{CoFeB}$ CoFeB plotted as a function of Bi$_{1-x}$Sb$_{x}$ thickness $t_\textrm{BiSb}$ (b,e), Sb concentration $x$ (c,f) and the measurement temperature $T$ (d,g).}
\label{fig:R_AHE_H_k}
\end{center}
\end{figure}

The Hall resistance $R_{\rm{xy}}$ of the heterostructures are measured using a dc current of 20 $\mu\rm{A}$. An exemplary loop of $R_{\rm{xy}}$ vs. $H_z$ is displayed in Fig.~\ref{fig:R_AHE_H_k}(a). The slope of $R_{\rm{xy}}$ at large magnetic field is attributed to the ordinary Hall effect of BiSb. Data at large magnetic field is fitted using a linear function. The anomalous Hall resistance $R_{\rm{AHE}}$ is obtained from the $y$-axis intercept of the fitted linear function shown by the red dotted lines\cite{LauJJAP2017_Harmonic_IP}. The low field data are also fitted with a linear function, as shown by the blue dotted line. The out-of-plane anisotropy field $H_{\rm{K}}$ is determined by the $x$-intercept of the two linear functions: see Fig.~\ref{fig:R_AHE_H_k}(a) for an illustration to obtain $H_{\rm{K}}$ from the loop.

Figure~\ref{fig:R_AHE_H_k}(b) and (e) show the layer thickness dependence of $R_{\rm{AHE}}$ and $H_{\rm{K}}$, respectively. As evident, $|R_{\rm{AHE}}|$ decreases with increasing Bi$_{1-x}$Sb$_{x}$ thickness regardless of the size of $t_{\textrm{CoFeB}}$ (Fig.~\ref{fig:R_AHE_H_k}(b)), which is attributed to current shunting into the Bi$_{1-x}$Sb$_{x}$ layer. Figure~\ref{fig:R_AHE_H_k}(e) shows that $H_{\rm{K}}$ is nearly independent of the Bi$_{1-x}$Sb$_{x}$ layer thickness. $H_{\rm{K}}$ increases with decreasing CoFeB layer thickness, which is consistent with the presence of a perpendicular magnetic anisotropy at the CoFeB/MgO  interface\cite{Ikeda_2010}. The Sb concentration ($x$) dependence of $R_{\rm{AHE}}$ and $H_{\rm{K}}$ are displayed in Fig.~\ref{fig:R_AHE_H_k}(c) and \ref{fig:R_AHE_H_k}(f), respectively. We find $|R_{\rm{AHE}}|$ decreases with increasing Sb concentration, which is predominantly due to larger current shunting into the Bi$_{1-x}$Sb$_{x}$ layer as the layer conductivity increases with increasing $x$ (see Fig. 3(a)). $H_{\rm{K}}$ tends to increase with increasing $x$. Although the underlying mechanism is not clear, these results indicate that the perpendicular magnetic anisotropy, predominantly defined at the CoFeB/MgO interface, is influenced by the underlayer\cite{Sinha2013apl}. The temperature dependence of $R_{\rm{AHE}}$ and $H_{\rm{K}}$ for heterostructures with 10 Bi$_{\rm{1-x}}$Sb$_{\rm{x}}$/2 CoFeB are shown in Fig.~\ref{fig:R_AHE_H_k}(d) and (g), respectively. For all $x$ studied, both $|R_{\rm{AHE}}|$ and $H_{\rm{K}}$ tend to increase with decreasing temperature. Changes in $R_{\rm{AHE}}$ are mainly attributed to the different temperature dependence of the conductivity of Bi$_{1-x}$Sb$_{x}$ and CoFeB. Whereas the conductivity of CoFeB is practically independent of the temperature, that of Bi$_{1-x}$Sb$_{x}$ decreases with decreasing temperature, resulting in redistribution of the current within the two layers on varying the temperature. The increase of $H_{\rm{K}}$ with temperature is unclear and requires further investigation.

\section{F\MakeLowercase{irst harmonic} H\MakeLowercase{all resistance of} B\MakeLowercase{i}S\MakeLowercase{b}/C\MakeLowercase{o}F\MakeLowercase{e}B}
\label{sec:harmonic}
\begin{figure}
\begin{center}
\includegraphics[width=5in]{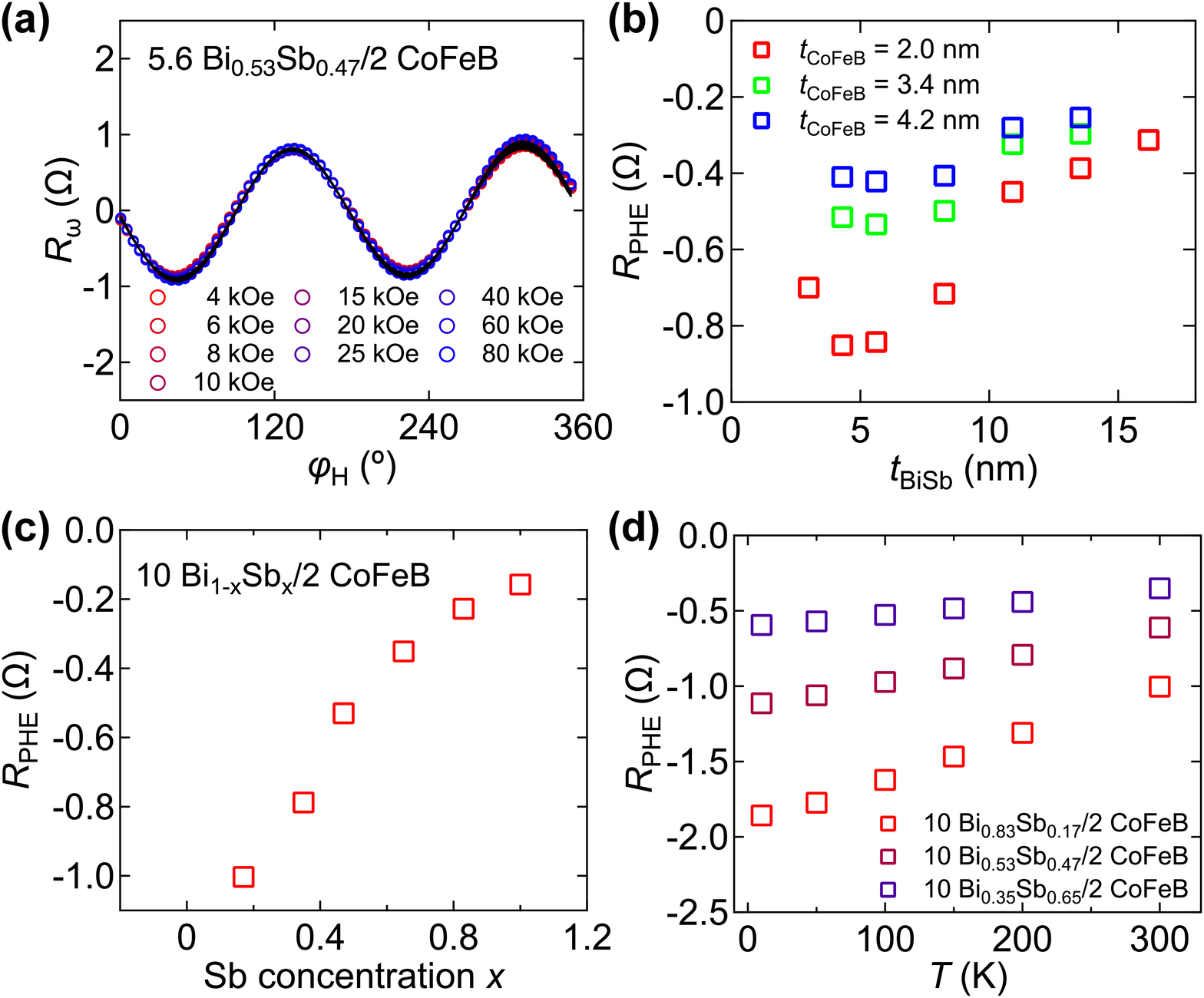}\\
\caption{\textbf{First harmonic Hall resistance.} (a) Field angle $\varphi_{\rm{H}}$ dependence of the first harmonic Hall resistance $R_{1\omega}$ of a heterostructure with 5.6 Bi$_{0.53}$Sb$_{0.47}$/2 CoFeB. Different symbols represent results obtained from various strengths of in-plane magnetic field. The black lines show the fitting with Eq.~(\ref{eq:R_PHE}). (b-d) Planar Hall resistance $R_\textrm{PHE}$ plotted against Bi$_{1-x}$Sb$_{x}$ thickness $t_\textrm{BiSb}$ (b), Sb concentration $x$ (c) and the measurement temperature $T$ (d), respectively.}
\label{fig:R_PHE}
\end{center}
\end{figure}

%Fig.~\ref{fig:R_PHE}(a) shows the angular dependence of $R_{1\omega}$ for 5.6 Bi$_{0.53}$Sb$_{0.47}$/2 CoFeB bilayer at different magnetic field.
Exemplary results of the first harmonic Hall resistance $R_{1\omega}$ plotted against the angle $\varphi_\textrm{H}$ (the angle between the external magnetic field applied along the film plane and the $x$ axis) are shown in Fig.~\ref{fig:R_PHE}(a). The results can be fitted using the following function:
\begin{equation}
\centering
\label{eq:R_PHE}
R_{1\omega}=R_{\rm{PHE}}\sin{2\varphi_{\rm{H}}}+{\zeta}R_{\rm{AHE}}\cos{\varphi_{\rm{H}}},
\end{equation}
where $\zeta$ is a constant that is introduced to take into account an unintentional misalignment between the external magnetic field and the film plane. We use Eq.~(\ref{eq:R_PHE}) to obtain the planar Hall resistance $R_{\rm{PHE}}$. We find little change in $R_{1\omega}$ on varying the magnetic field, suggesting good sample alignment and weak external field dependence of $R_{\rm{PHE}}$. We therefore take the average $R_{\rm{PHE}}$ obtained under different magnetic fields.
% to estimate $H_{\rm{FL}}+H_{\rm{Oe}}$.
%The planar Hall resistance ($R_{\rm{PHE}}$) used for the estimation of the sum of the field-like spin-orbit torque effective field $H_{\rm{FL}}$ and Oersted field  $H_{\rm{Oe}}$ is extracted from $R_{1\omega}$. Considering a possible sample misalignment resulting in a $R_{\rm{AHE}}$ contribution to $R_{1\omega}$, the angular dependence of $R_{1\omega}$ should be written as

Figure~\ref{fig:R_PHE}(b) illustrates the Bi$_{1-x}$Sb$_{x}$ and CoFeB layer thickness dependence of $R_{\rm{PHE}}$. Note that in heavy metal/CoFeB bilayers $R_{\rm{PHE}}$ contains contribution from the spin Hall magnetoresistance (SMR)\cite{Kim2016PRL,Cho_2015}. We find a peak in $| R_{\rm{PHE}} |$ at $t_{\rm{BiSb}}\sim$ 5 nm, which is approximately twice of the Bi$_{1-x}$Sb$_{x}$ spin diffusion length ($\sim$ 2.3 nm) determined in the main text. The position of the peak in $| R_{\rm{PHE}} |$ is consistent with the SMR theory\cite{Chen2013PRB}.
$R_{\rm{PHE}}$ for heterostructures with 10 Bi$_{\rm{1-x}}$Sb$_{\rm{x}}$/2 CoFeB are plotted as a function of Sb concentration ($x$) and temperature in Fig.~\ref{fig:R_PHE}(c) and \ref{fig:R_PHE}(d) respectively. We consider these changes are mainly due to variation of the spin Hall efficiency and current redistribution within the bilayers.
%\section{T\MakeLowercase{hermoelectric effects}}
%\label{sec:TEE}
%Contributions from the thermoelectric effects $V_{\rm{const}}$ and $V_{\rm{ONE}}$ on $R_{2\omega}$ are plotted as a function of $I_{\rm{CoFeB}}^2$ and $I_{\rm{BiSb}}^2$ in Fig.~\ref{fig:TEE}(a) and (b), respectively. Here data from heterostructures with $t_\textrm{BiSb}$ Bi$_{0.53}$Sb$_{0.47}$/$t_\textrm{CoFeB}$ CoFeB are shown. %Data were measured using a constant current $I$=1.5 mA.
%$I_{\rm{CoFeB}}$ and $I_{\rm{BiSb}}$ denote the current flowing in CoFeB and BiSb layer, respectively, estimated using a parallel resistance model. We find that $V_{\rm{const}}$ ($V_{\rm{ONE}}$) is proportional to the power dissipation, i.e. the square of the current flowing within the CoFeB (BiSb) layer, confirming the thermoelectric origin of these $R_{2\omega}$ contributions.

\section{T\MakeLowercase{wo-band model analysis of} B\MakeLowercase{i}S\MakeLowercase{b}}
\label{sec:twoband}
\begin{figure}
\begin{center}
\includegraphics[width=5in]{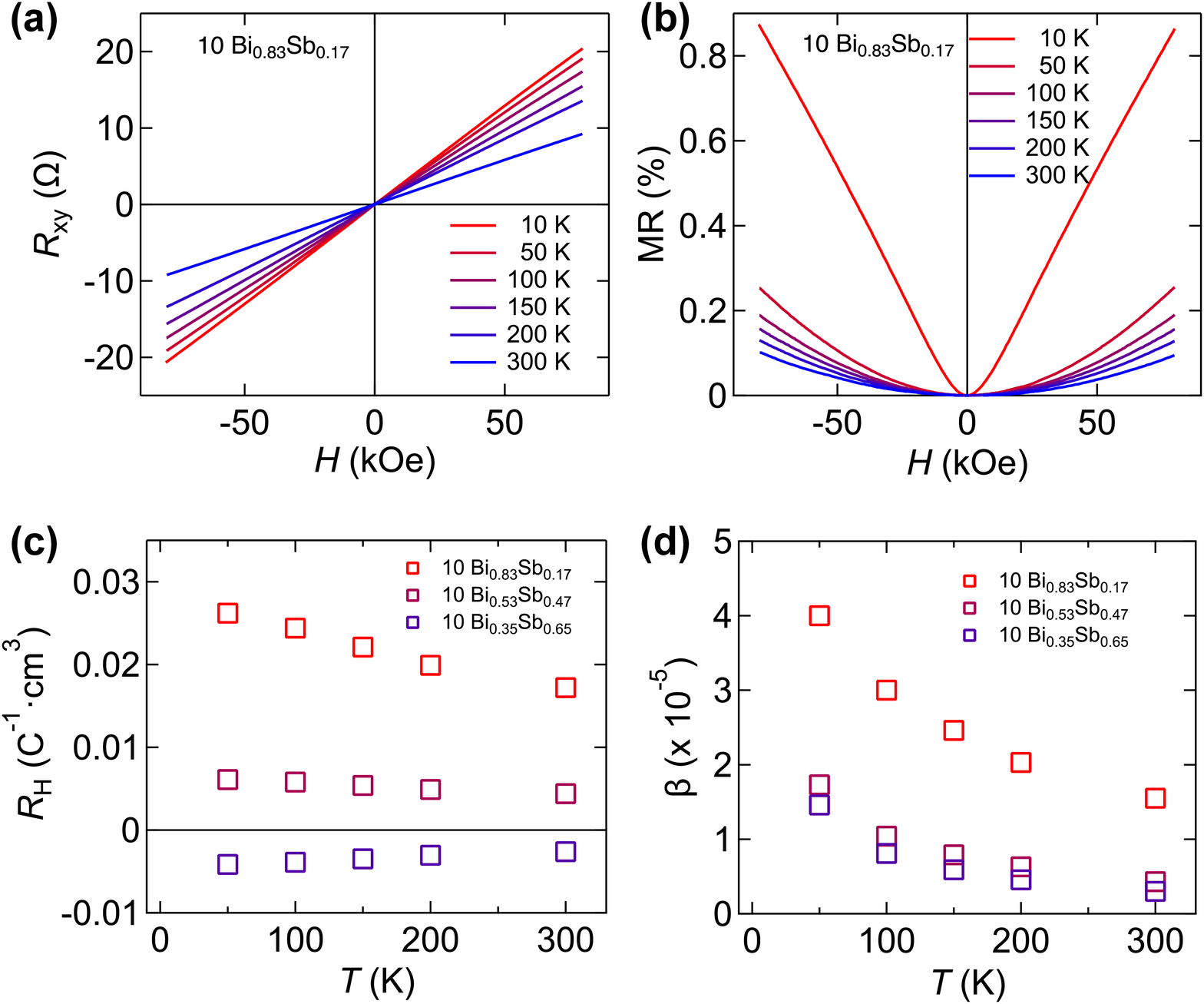}\\
\caption{\textbf{Temperature dependence of magneto-transport properties of Bi$_{1-x}$Sb$_{x}$.} (a-b) Temperature dependence of $R_\textrm{xy}$ (a) and the magnetoresistance (MR) (b) plotted against external field $H_\textrm{ext}$ along $z$ for a heterostructure with10 Bi$_{0.83}$Sb$_{0.17}$ (without the CoFeB layer). (c,d) Hall coefficient $R_\textrm{H}$ (c) and MR coefficient $\beta$ (d) for Bi$_{1-x}$Sb$_{x}$ heterostructures  (without the CoFeB layer) of different compositions plotted as a function of temperature.}
\label{fig:beta_R_H}
\end{center}
\end{figure}

Carrier concentration and mobility of Bi$_{1-x}$Sb$_{x}$ are estimated based on a classical two-carrier model, \textit{i.e.} both electrons and holes contribute to the transport. Experimental inputs of the model are: Hall coefficient $R_{\rm{H}}$ (given by $R_{\rm{H}}=R_{xy}t_{\rm{BiSb}}/H_{z}$, where $H_{z}$ represents $H_{\textrm{ext}}$ along the $z$-axis), longitudinal conductivity at zero field $\sigma_{\rm{BiSb}}$ and the quadratic component of the transverse magnetoresistance (MR) with $H_z$. The equations read\cite{AliNature2014}:
\begin{equation}
\centering
\label{eq:RH_origin}
R_{\rm{H}}=\frac{(n_{e}\mu_{e}^{2}-n_{h}\mu_{h}^{2})+(n_{h}-n_{e})\mu_{h}^{2}\mu_{e}^{2}H_{z}^2}{e\left(n_{h}\mu_{h}+n_{e}\mu_{e}\right)^{2}+(n_{h}-n_{e})^2\mu_{h}^{2}\mu_{e}^{2}H_{z}^2},
\end{equation}
\begin{equation}
\centering
\label{eq:sigma_origin}
\sigma_{\rm{BiSb}}=e\left(n_{h}\mu_{h}+n_{e}\mu_{e}\right),
\end{equation}
\begin{equation}
\centering
\label{eq:MR_origin}
\rm{MR}=\frac{\rho_{\rm{BiSb}}\left(\it{H}_{z}\right)-\rho_{\rm{BiSb}}\left(\it{H}_{z}=0\right)}{\rho_{\rm{BiSb}}\left(\it{H}_{z}=0\right)}=\frac{n_{h}\mu_{h}n_{h}\mu_{h}\left(\mu_{h}+\mu_{e}\right)^{2}\it{H}_{z}^2}{\left(n_{h}\mu_{h}+n_{e}\mu_{e}\right)^2+\left(n_{h}-n_{e}\right)^2\left(\mu_{h}\mu_{e}\right)^2\it{H}_{z}^2}.
\end{equation}
Here, $n_{h}$($n_{e}$) and $\mu_{h}$($\mu_{e}$) denote the carrier density and the mobility of hole(electron), respectively. Since Bi and Sb have the same number of \textit{p} valence electrons, we assume $n_{h}\approx n_{e}\equiv n$ for all $x$ of Bi$_{\rm{1-x}}$Sb$_{\rm{x}}$. With such carrier compensation, Eqs.~(\ref{eq:RH_origin}), (\ref{eq:RH_origin}) and (\ref{eq:RH_origin}) can be reduced to:
%\ref$R_{\rm{H}}$, $\sigma_{\rm{BiSb}}$ and MR
\begin{equation}
\centering
\label{eq:RH}
R_{\rm{H}}=\frac{ne\left(\mu_{e}^{2}-\mu_{h}^{2}\right)}{\sigma_{\rm{BiSb}}^{2}},
\end{equation}
\begin{equation}
\centering
\label{eq:sigma}
\sigma_{\rm{BiSb}}=ne\left(\mu_{h}+\mu_{e}\right),
\end{equation}
\begin{equation}
\centering
\label{eq:MR_origin}
\rm{MR}=\mu_{e}\mu_{h}\it{H}_{z}^{2}=\beta \it{H}_{z}^{2}.
\end{equation}
Based on these equations, $n$ and mobilities $\mu_{h}$, $\mu_{e}$ can be estimated from experimental results. As an example, the temperature dependences of $R_\textrm{xy}$ and MR for 10 Bi$_{0.83}$Sb$_{0.17}$ grown on Ta seed layer are shown in Fig.~\ref{fig:beta_R_H}(a) and \ref{fig:beta_R_H}(b). $R_{\rm{H}}$ and $\beta \equiv \mu_{e}\mu_{h}$ extracted from the plots are shown in Fig.~\ref{fig:beta_R_H}(c) and \ref{fig:beta_R_H}(d), respectively. The large quasi-linear MR at 10 K is mainly attributed to the weak anti-localization of Bi$_{1-x}$Sb$_{x}$ exhibiting strong spin orbit coupling. We therefore limit our analysis of carrier density and mobility to temperature ranging from 50 K to 300 K. The estimated $n$, $\mu_{h}$ and $\mu_{e}$ are shown in Fig.~5 of the manuscript.

\section{E\MakeLowercase{valuation of the} SOT \MakeLowercase{analysis protocol}}
\label{sec:Pt}
\begin{figure}
\begin{center}
\includegraphics[width=6.5in]{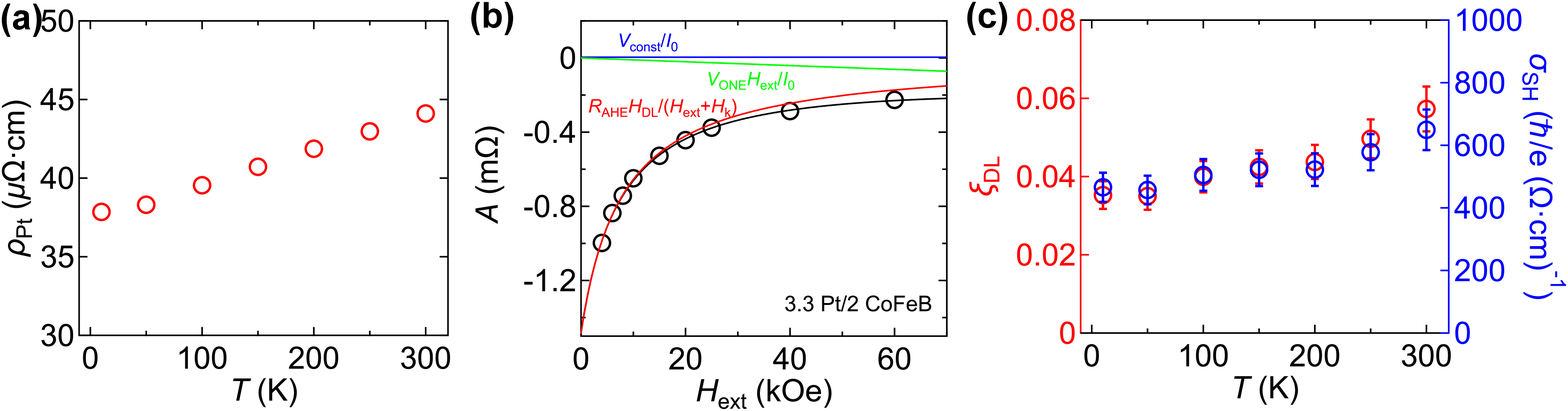}\\
\caption{\textbf{SOT measurements of a standard sample: Pt/CoFeB.} (a-c) The film structure is Sub./1 Ta/3.3 Pt/2 CoFeB/2 MgO/1 Ta (thicknesses in nm). (a) Temperature dependence of the Pt layer resistivity $\rho_{\rm{Pt}}$. (b) $H_{\rm{ext}}$ dependence of fitting parameter $A$ obtained at measurements carried out at room temperature. (c) Temperature dependence of the damping-like spin Hall efficiency $\xi_{\rm{DL}}$ and the spin Hall conductivity $\sigma_{\rm{SH}}$.}
\label{fig:Pt}
\end{center}
\end{figure}

To verify the analysis protocol used here, the damping-like and field-like SOTs for a standard sample, Sub./1 Ta/3.3 Pt/2 CoFeB/2 MgO/1 Ta (thicknesses in nm) heterostructure, was measured. 
%We compare the charge to spin conversion efficiency of Pt with that of Bi$_{1-x}$Sb$_x$ alloys. 
Figure~\ref{fig:Pt}(a) shows the resistivity of Pt ($\rho_{\rm{Pt}}$) as a function of temperature. A parallel circuit model is employed to estimate the resistivity of Pt; we assume the only other conducting layer is CoFeB, which has a resistivity of $\rho_{\rm{CoFeB}}$$\sim$140 $\mu\Omega$cm. We find $\rho_{\rm{Pt}}$ is $\sim$44 $\mu\Omega\rm{cm}$ at room temperature and decreases to $\sim$38 $\mu\Omega\rm{cm}$ at 5 K, reflecting the metallic transport property of Pt (the small residual-resistance ratio indicates large amount of impurity in Pt). Parameter $A$ at 300K, which is defined in Eq.~(2) of the main text, is plotted as a function of $H_{\rm{ext}}$ in Fig.~\ref{fig:Pt}(b). The colored lines represent contributions from different components. Thermoelectric signals in Pt/CoFeB are sufficiently small compared to the signal due to SOT. Figure~\ref{fig:Pt}(c) shows the damping-like spin Hall efficiency $\xi_{\rm{DL}}$ and the spin Hall conductivity $\sigma_{\rm{SH}}$ plotted as a function of temperature. We find $\xi_{\rm{DL}}$$\sim$0.06 at room temperature, corresponding to $\sigma_\textrm{SH}$$\sim$ 650 $(\hbar/e)\Omega^{-1}$cm$^{-1}$. $\xi_{\rm{DL}}$ and $\sigma_{\rm{SH}}$ slightly increase with increasing temperature, consistent with previous results obtained in similar heterostructures (Pt/Py) evaluated using the spin-torque ferromagnetic resonance (ST-FMR) technique\cite{Wang2014APL} and lateral spin valve measurements\cite{SagastaPRB2016_Scaling_SHA_Pt}.

\section{T\MakeLowercase{he efficiency of} B\MakeLowercase{i}S\MakeLowercase{b} SOT}
\label{sec:Jc}
Although BiSb exhibit a large spin Hall efficiency, its resistivity is larger than typical metals, including the ferromagnetic metals that are used when forming SOT devices. Here we estimate the critical current needed to control the magnetization direction of the adjacent ferromagnet using BiSb and compare it with other materials. We consider a bilayer composed of a ferromagnetic metal (FM) layer and a non-magnetic metal (NM) layer, i.e. the latter generates spin current via the spin Hall effect. The critical current density is defined as the NM layer charge current density $j_{\rm{NM}}$ required to generate sufficient damping-like SOT to induce switching of the FM layer magnetization.
%The length and width of the device is assumed to be same for FM and spin Hall source layers. 
In the macrospin limit, $j_{\rm{NM}}$ takes the form\cite{lee2014apl}: 
\begin{equation}
\centering
\label{eq:J_NM}
j_{\rm{NM}}=\frac{e H_{\textrm{K,eff}} M_{\textrm{s}} t_{\rm{FM}}}{2 \hbar \xi_{\rm{DL}}} f(h_x),
\end{equation}
where $H_{\textrm{K,eff}} $ is the effective anisotropy field, $M_{\textrm{s}}$ is the saturation magnetization of the FM layer, and $f(h_x)$ is a function of $h_x \equiv \frac{H_x}{H_{\textrm{K,eff}}}$. $H_x$ is the external in-plane field $H_x$, $e$ and $\hbar$ are the electric charge and the reduced Planck constant, respectively. Here we have assumed a perpendicularly magnetized system to estimate the switching current; however, the only factor that is relevant in the following discussion is the spin Hall efficiency $\xi_{\textrm{DL}}$. 
For simplicity, we assume the properties of the FM layer do not depend on the NM layer. 
%Thus the $H_{c}$ and $M_\textrm{eff}$ are not relevant in the following discussion.)
The resistivity of the NM and FM layers are defined as $\rho_{\rm{NM}}$ and $\rho_{\rm{FM}}$, respectively. Assuming a parallel circuit, the overall current that flows in the bilayer when $j_{\textrm{NM}}$ is applied to the NM layer is given as
\begin{equation}
\centering
\label{eq:J_c}
I_{\textrm{c}}=j_{\rm{NM}} w \frac{\rho_{\rm{NM}}t_{\rm{FM}}+\rho_{\rm{FM}}d_{\rm{NM}}}{ \rho_{\rm{FM}}},
\end{equation}
where $w$ is the width of the device, which we assume is the same for both NM and FM layers. To compare the efficiency of BiSb with other materials system, we substitute typical values of the relevant parameters and estimate $I_c$. We compare BiSb/CoFeB with W/CoFeB\cite{PaiAPL2012_SHE_W}, the latter being the prototype of SOT-MRAM\cite{garello2018ieee}. The parameters assumed are: $t_{\rm{CoFeB}}=1$ nm and $\rho_{\rm{CoFeB}}=150$ $\mu\Omega$cm
for FM=CoFeB, $d_{\rm{W}}=5$ nm, $\rho_{\rm{W}}=120$ $\mu\Omega$cm, $\xi_{\textrm{DL,W}}$=0.3 for NM=W, and $d_{\rm{BiSb}}=10$ nm, $\rho_{\rm{BiSb}}=1000$ $\mu\Omega$cm, $\xi_{\textrm{DL,BiSb}}$=1.2 for NM=BiSb. The ratio of the critical current becomes
\begin{equation}
\centering
\label{eq:Jc_ratio}
\frac{I_{\textrm{c,W}}}{I_{\textrm{c,BiSb}}}=\frac{\xi_{\textrm{DL,BiSb}}}{\xi_{\textrm{DL,W}}}\frac{\rho_{\rm{W}}t_{\rm{CoFeB}}+\rho_{\rm{CoFeB}}d_{\rm{W}}}{\rho_{\rm{BiSb}}t_{\rm{CoFeB}}+\rho_{\rm{CoFeB}}d_{\rm{BiSb}}} \sim 1.4.
\end{equation}
Thus we find the large spin Hall efficiency of BiSb is beneficial for SOT-MRAM even for its large resistivity: the current can be reduced by $\sim$40\% compared to the state of art materials system, W/CoFeB.

\clearpage

%\begin{figure}
%\begin{center}
%\includegraphics[width=5in]{SM_TEE_thickness}\\
%\caption{\textbf{Thermoelectric contributions to the second harmonic Hall resistance.} (a,b) $V_{\rm{const}}$ (a) and $V_{\rm{ONE}}$ (b) against the square of the current flowing in CoFeB and Bi$_{0.53}$Sb$_{0.47}$ layer, respectively. Results are from heterostructures with textcolor{red}{(insert thickness of BiSb)}$t_\textrm{BiSb}$ Bi$_{0.53}$Sb$_{0.47}$/$t_\textrm{CoFeB}$ CoFeB.}
%\label{fig:TEE}
%\end{center}
%\end{figure}

\clearpage
\bibliography{BiSb_ref_081919}

\end{document}